\documentclass[12pt]{article}
\usepackage{amsfonts}

\usepackage{latexsym}
\usepackage{amssymb}
\usepackage{epsfig}
\usepackage{amsmath}
\usepackage{amsthm}
\usepackage[mathscr]{eucal}
\usepackage{subfigure}
\usepackage{graphicx}

\usepackage[linesnumbered,ruled,vlined]{algorithm2e}

\SetKwInOut{Input}{Input}
\SetKwInOut{Output}{Output\,}
\SetKwInOut{Data}{Data}
\SetKwProg{SequentialDecode}{SequentialDecode}{}{EndSequentialDecode}
\SetKwProg{SeqDecodeMod}{SeqDecodeMod}{}{EndSeqDecodeMod}
\SetKwProg{CorrectDeletion}{CorrectDeletion}{}{EndCorrectDeletion}
\SetKwProg{CorrectFlip}{CorrectFlip}{}{EndCorrectFlip}
\SetKwProg{CorrectErasure}{CorrectErasure}{}{EndCorrectErasure}
\SetKwProg{BurstDecode}{BurstDecode}{}{EndBurstDecode}
\SetKwProg{OrdDecode}{OrdDecode}{}{EndOrdDecode}
\SetKwProg{MultDecode}{MultDecode}{}{EndMultDecode}
\SetKwProg{MultDelErrDecode}{MultDelErrDecode}{}{EndMultDelErrDecode}

\makeatletter
\def\BState{\State\hskip-\ALG@thistlm}
\makeatother

\newtheorem{Theorem}{Theorem}

\newtheorem{Lemma}{Lemma}[section]
\newtheorem{Corollary}{Corollary}

\renewcommand{\qed}{\hfill{\ \ \rule{2mm}{2mm}} \vspace{0.2in}}

\newcommand{\ind}{1\hspace{-2.3mm}{1}}

\begin{document}

\title{Construction and redundancy of codes for correcting deletable errors}
\author{ \textbf{Ghurumuruhan Ganesan}
\thanks{E-Mail: \texttt{gganesan82@gmail.com} } \\
\ \\
New York University, Abu Dhabi}
\date{}
\maketitle

\begin{abstract}
Consider a binary word being transmitted through a communication channel that introduces deletable errors where each bit of the word is either retained, flipped, erased or deleted. The simplest code for correcting \emph{all} possible deletable error patterns of a fixed size is the repetition code whose redundancy grows linearly with the code length. In this paper, we relax this condition and construct codes capable of correcting \emph{nearly} all deletable error patterns of a fixed size, with redundancy growing as a logarithm of the word length.

\vspace{0.1in} \noindent \textbf{Key words:} Deletable errors, low redundancy codes.

\vspace{0.1in} \noindent \textbf{AMS 2000 Subject Classification:} Primary:
60J10, 60K35; Secondary: 60C05, 62E10, 90B15, 91D30.
\end{abstract}

\bigskip

\setcounter{equation}{0}
\renewcommand\theequation{\thesection.\arabic{equation}}
\section{Introduction} \label{intro}
Consider a binary word being transmitted through a communication channel that introduces errors in the individual bits. If the errors introduced are nondeletable errors like erasures or bit flippings (i.e., a one is converted to a zero and vice versa), then it is possible to use linear error correcting codes to correct any desired number of errors with low redundancy.

In case the errors introduced include deletions, the above codes are not directly applicable since there are synchronization problems. For the case of pure deletion errors, we briefly review related literature. If a single deletion corrupts a word, then Levenshtein (1965) showed that it is possible to use Varshamov-Tenengolts codes (VT codes) (Varshamov and Tenengolts (1965)) for recovering the original word. Since then extensive literature has appeared on the construction of codes capable of correcting multiple deletions with varying constraints. Helberg et al (2002) described codes for correcting arbitrary number of deletions and later Abdel-Ghaffer et al (2012) provided a theoretical analysis of the deletion correction capability of Helberg's codes. Helberg's codes have a redundancy that grows linearly with the word length. Brasniek et al (2016) used hashing techniques to construct low redundancy codes for correcting multiple deletions but with the caveat that the codewords belonged to a set of strings rich in certain predetermined patterns. Recently, Schoeny et al (2017) proposed a class of shifted VT codes to deal with burst deletions of fixed length. For a survey of literature on deletion channels, see Mitzenmacher (2009).




Random codes for deletions have also been studied before. From a communication complexity perspective, Orlitsky~(1993) obtained bounds for file synchronization via deletion correction  with differing constraints on the number of rounds communication allowed. In a related work, Venkatramanan et al (2015) focused on developing bidirectional interactive algorithms with low information exchange. Recently, Hanna and Rouayheb (2017) proposed Guess and Check codes that map blocks of codewords to symbols in higher fields and used Reed-Solomon erasure decoding to correct words corrupted by a fixed number of deletions distributed randomly across the word.

Most of the above literature mainly focus on correcting either pure deletions or purely nondeletable errors like erasures and flippings. In this paper, we seek codes with low redundancy that are capable of correcting deletable errors that could be either erasure, flipping or deletion. Our main results (see discussion following Corollary~\ref{cor_11}) state that there are codes with redundancy that grows logarithmically in the code length, capable of correcting \emph{nearly} all possible deletable error patterns of a fixed size.


\subsection*{System Description}
A word of length~\(n\) is a vector~\(\mathbf{x} = (x_1,\ldots,x_n) \in \{0,1\}^n.\) A deletable error pattern of length~\(n\) is an element~\(\mathbf{g} = (\mathbf{e},\mathbf{ty})  = ((e_1,\ldots,e_n),(ty_1,\ldots,ty_n))\in \{0,1\}^{n} \times \{D,E,F\}^{n},\) with~\(D, E\) and~\(F\) denoting deletion, erasure and flipping, respectively. The word~\(\mathbf{y} = F_{\mathbf{g}}(\mathbf{x})\) obtained after~\(\mathbf{x}\) is corrupted by~\(\mathbf{g}\) is defined as follows: If~\(e_i = 1,\) then the bit~\(x_i\) is deleted, erased or flipped depending on whether~\(ty_i = D, E\) or~\(F,\) respectively. If~\(e_i = 0,\) no change occurs to the bit~\(x_i.\) Thus for example, if~\(n = 5,\mathbf{x} = (x_1,x_2,x_3,x_4,x_5)\)
and~\(\mathbf{g} = ((1,0,1,1,1),(F,E,D,F,E)),\) then~\[\mathbf{y} = F_{\mathbf{g}}(\mathbf{x}) = (1-x_1,x_2,1-x_4,\varepsilon),\] where~\(\varepsilon\) is the erasure symbol. 


For~\(r \geq 1,\) we define
\begin{equation}\label{en_def}
{\cal E}_n(r) := \left\{(\mathbf{e},\mathbf{ty}) : \sum_{i=1}^{n} e_i \leq r \right\}
\end{equation}
to be the set of all possible~\(n-\)length patterns containing at most~\(r\) deletable errors. We also let~\({\cal E}_n = \cup_{1 \leq r \leq n} {\cal E}_n(r)\) be the set of all possible error patterns.

An~\(n-\)length code of size~\(q\) is a collection of words~\(\{\mathbf{x}_1,\ldots,\mathbf{x}_q\} \subset \{0,1\}^{n}.\) Let~\({\cal F} \subseteq {\cal E}_n\) be any set of~\(n-\)length error patterns. A~\(n-\)length code~\({\cal C}\) is said to be capable of correcting all error patterns in~\({\cal F}\) if for any~\(\mathbf{x}_1 \neq \mathbf{x}_2 \in {\cal C}\) and any~\(\mathbf{g}_1,\mathbf{g}_2 \in {\cal F},\) we have~\(F_{\mathbf{g}_1}(\mathbf{x}_1) \neq F_{\mathbf{g}_2}(\mathbf{x}_2).\) We also say that~\({\cal C}\) is a~\({\cal F}-\)correcting~\(n-\)length code and define the redundancy of~\({\cal C}\) to be
\begin{equation}\label{red_def}
R({\cal C}) := n - \log\left(\#{\cal C}\right).
\end{equation}
Throughout all logarithms are to the base~\(2.\)

We have the following result regarding the minimum redundancy of codes capable of correcting deletable errors.
\begin{Theorem}\label{prop1} Let~\(\{t_n\}\) be any sequence such that~\(\frac{t_n\log{n}}{n} \longrightarrow 0\) as~\(n \rightarrow \infty.\) For all~\(n \geq 3,\) there exists a~\(n-\)length code~\({\cal C}_{del} = {\cal C}_{del}(n)\) with redundancy
\begin{equation}\label{red_tn}
n\left(1-\frac{1}{2t_n+1}\right) \leq R({\cal C}_{del}) \leq n\left(1-\frac{1}{2t_n+1}\right) + 1,
\end{equation}
that is capable of correcting up to~\(t_n\) deletable errors; i.e., all error patterns in~\({\cal E}_n(t_n).\) Conversely, if~\({\cal A}\) is any~\(n-\)length code capable of correcting up to~\(t_n\) errors, then
\begin{equation}\label{red_tn_conv}
R({\cal A}) \geq t_n \log\left(\frac{n}{t_n}\right) - 10t_n - \frac{2^{11}t_n^2}{n} - 1
\end{equation}
for all~\(n\) large. 
\end{Theorem}
The code~\({\cal C}_{del}\) in Proposition~\ref{prop1} is a repetition code and the encoding and decoding algorithms are described in Section~\ref{prelim}.

The redundancy of the code in Theorem~\ref{prop1} grows linearly with the code length~\(n.\) This is in part because of the strict condition that \emph{all} deletable errors affecting up to~\(t_n\) bits must be corrected.  We now relax this condition by stipulating that \emph{nearly all} error patterns affecting up to~\(t_n\) bits need to be corrected and construct codes with low redundancy for correcting
such patterns.

For~\(n \geq 1\) let~\({\cal C}\) be a~\(n-\)length code and~\(1 \leq t_n \leq n\) be any integer. Recall from discussion prior to~(\ref{en_def}) that~\({\cal E}_n(t_n)\) is the set of all possible error patterns containing at most~\(t_n\) deletable errors. We say that~\({\cal C}\) corrects at least a fraction~\(f \in (0,1)\) of patterns in~\({\cal E}_n(t_n)\) if there exists~\({\cal F} \subseteq {\cal E}_n(t_n)\) such that~\({\cal C}\) is~\({\cal F}-\)correcting and~\(\#{\cal F} \geq f \cdot \#{\cal E}_n(t_n).\)


We have the following result regarding existence and redundancy of codes capable of correcting nearly all deletable errors.
\begin{Theorem}\label{thm_12} Let~\(\{t_n\}\) and~\(\{\omega_n\}\) be positive sequences such that~\(1 \leq t_n \leq n\) and~\(\omega_n \geq 6\) for all~\(n\) and
\begin{equation}\label{omega_cond}
\frac{\omega_n t^3_n}{n} \longrightarrow 0
\end{equation}
as~\(n \rightarrow \infty.\) For all~\(n\) large, there is a~\(n-\)length code~\({\cal C}_{frac} = {\cal C}_{frac}(n)\) with
redundancy
\begin{equation}\label{red_upp}
R({\cal C}_{frac}) \leq \omega_n t_n^2 \log\left(\frac{2n}{\omega_n t_n^2}\right), 
\end{equation}
that is capable of correcting at least a fraction~\(1-\frac{42}{\omega_n}\) of the patterns in~\({\cal E}_n(t_n).\)
\end{Theorem}
Since~\(\frac{\omega_n t_n^2}{n} \leq \frac{\omega_n t_n^3}{n} \longrightarrow 0\) (see~\ref{omega_cond})), we get from~(\ref{red_upp}) that~\(\frac{R({\cal C}_{frac})}{n} \longrightarrow 0\) as~\(n \rightarrow \infty.\) Thus the redundancy of the code~\({\cal C}_{frac}\) is much smaller than~\(n\) for all~\(n\) large. 

As a consequence of Theorem~\ref{thm_12}, we have the following Corollary.
\begin{Corollary}\label{cor_11} Let~\(\{t_n\}\) be any sequence such that~\(\frac{t^3_n}{n} \longrightarrow 0\) as~\(n \rightarrow \infty.\) For every integer~\(K \geq 2,\) there exists an integer~\(N(K)\) such that if~\(n \geq N(K),\) then there is a~\(n-\)length code~\({\cal C}^{(K)}_{frac}= {\cal C}^{(K)}_{frac}(n)\) with redundancy
\begin{equation}\label{red_upp2}
R({\cal C}^{(K)}_{frac}) \leq K t_n^2 \log\left(\frac{2n}{Kt_n^2}\right), 
\end{equation}
that is capable of correcting at least a fraction~\(1-\frac{42}{K}\) of the patterns in~\({\cal E}_n(t_n).\)
\end{Corollary}
Corollary~\ref{cor_11} follows from Theorem~\ref{thm_12} by fixing integer~\(K \geq 1\) and setting~\(\omega_n = K\) for all~\(n.\)
If the number of deletable errors~\(t_n = t\) is a constant, then the code~\({\cal C}^{(K)}_{frac}, K \geq 2\) described in Theorem~\ref{thm_12} has redundancy of at most~\(Kt^2 \log{n}\) for all~\(n\) large and is capable of correcting at least a fraction of~\(1-\frac{1}{K}\) of the error patterns affecting up to~\(t\) bits. For comparison, we recall from Theorem~\ref{prop1} that if a code~\({\cal A}\) is capable of correcting up to~\(t\) deletable errors, then its redundancy necessarily is at least of the order of~\(t\log{n}.\)

\subsection*{Special error patterns}
In this subsection, we consider correction of various special deletable error patterns.

\subsubsection*{\(P-\)far deletable error patterns}
Let~\(n \geq P \geq 2\) be any integers. If~\(\mathbf{g} =(\mathbf{e},\mathbf{ty}), \mathbf{e} = (e_1,\ldots,e_n) \in \{0,1\}^{n}\) is an error pattern such that~\(|i-j| \geq P\) for any two non zero~\(e_i,e_j,\) we then say that~\(\mathbf{g}\) is a (\(n-\)length) \(P-\)far deletable error pattern.
Letting
\begin{equation}\label{del_def_main}
\delta(P) := \frac{P+1}{2^{P-1}}
\end{equation}
so that~\(\delta(P) \longrightarrow 0\) as~\(P \rightarrow \infty,\) we have the following result regarding codes capable of correcting~\(3P-\)far deletable
error patterns.
\begin{Theorem}\label{thm_11} For all integers~\(n \geq P \geq 2\) there exists a \(n-\)length code~\({\cal C}_{far} = {\cal C}_{far}(n,P)\) with redundancy
\begin{equation}\label{red_upp44}
R({\cal C}_{far}) \leq \left(\frac{n}{P}-1\right)\log\left(\frac{P+1}{1-\delta(P)}\right) + \log{P} + 1, 
\end{equation}
that is capable of correcting all~\(3P-\)far deletable error patterns. Conversely, if~\({\cal A}\) is any \(n-\)length code capable
of correcting all~\(3P-\)far deletable error patterns, then
\begin{equation}\label{min_red2}
R({\cal A}) \geq  \frac{n}{2^{11}(3P+6)} - 2  
\end{equation}
for all large~\(n.\)

Suppose~\(\{P_n\}\) is a sequence satisfying
\begin{equation}\label{pn_cond}
\frac{P_n}{\sqrt{n\log{n}}} \longrightarrow \infty
\end{equation}
as~\(n \rightarrow \infty.\) If~\({\cal B}\) is any \(n-\)length code capable
of correcting all~\(3P_n-\)far deletable error patterns, then
\begin{equation}\label{min_red22}
R({\cal B}) \geq  \left(\frac{n}{6P_n}-1\right)\log\left(\frac{3P_n}{64}\right) - 2 
\end{equation}
for all large~\(n.\)
\end{Theorem}


We now consider the case of deletable error bursts. For a deletable error pattern~\(\mathbf{g} = (\mathbf{e},\mathbf{ty}), \mathbf{e} = (e_1,\ldots,e_n),\) let \(S(\mathbf{g}) = \{i : e_i = 1\}.\) We say that~\(\mathbf{g}\) is a burst deletable error pattern of length at most~\(b\) if
\begin{equation}\label{err_bur}
\max\{j : j \in S(\mathbf{g})\} - \min\{j : j \in S(\mathbf{g})\} \leq b.
\end{equation}
Our definition includes the case where the errors are not necessarily consecutive but all occur within a block of size~\(b.\) We have the following result.
\begin{Theorem}\label{thm3} For all~\(n \geq 3,\) there is a~\(n-\)length code~\({\cal C}_{bur}\) with redundancy
\[n\left(1-\frac{1}{2b+1}\right) \leq R({\cal C}_{bur}) \leq n\left(1-\frac{1}{2b+1}\right) + 1,\] that is capable of correcting all burst deletable error patterns of length at most~\(b.\) Conversely, if~\({\cal A}\) is any~\(n-\)length code capable of correcting all deletable error bursts of length at most~\(b,\) then the redundancy
\begin{equation}\label{min_red}
R({\cal A}) \geq  \log{n} -(b+5) - \log(b(b+4)), 
\end{equation}
for all~\(n\) large.
\end{Theorem}

The paper is organized as follows: In Section~\ref{prelim}, we present all the encoding and decoding algorithms needed for the proofs of  the main theorems. In~Section~\ref{pf2}, we prove the main Theorems apart from the redundancy lower bounds. Finally, in Section~\ref{pf_red_min}, we prove all the redundancy lower bounds in the above Theorems.






\setcounter{equation}{0}
\renewcommand\theequation{\thesection.\arabic{equation}}
\section{Algorithms}\label{prelim}
In this section, we present all the encoding and decoding algorithms needed for the proof of Theorem~\ref{thm_12}. We begin with a brief review of codes capable of correcting a single deletable error. 

\subsection{Single deletable error correction}
For integers~\(n \geq 3\) and~\(0 \leq a \leq n,\) define the Varshamov-Tenengolts (VT) code~\(VT_{a}(n)\)
\begin{equation}\label{vt_def}
VT_{a}(n) := \left\{\mathbf{x} = (x_1,\ldots,x_n) \in \{0,1\}^{n} : \sum_{i=1}^{n} i \cdot x_i \equiv a \mod{n+1}\right\}.
\end{equation}
From Levenshtein (1965), we know that the VT codes are capable of correcting a single deletion. The code~\(VT_a(n)\) is also capable of correcting a single deletable error and for completeness, we present the correction algorithm below.

\emph{Correcting an erasure}: Suppose~\(\mathbf{y} = F_{\mathbf{g}}(\mathbf{x}) = (y_1,\ldots,y_n)\) is obtained after a single erasure in the word~\(\mathbf{x}\) so that bit~\(y_k = \varepsilon\) (the erasure symbol) for some~\(1 \leq k \leq n.\) We then correct the erasure by computing the checksum
\begin{equation}\label{er_chk_sum22}
CS_{er} := \sum_{i=1, i \neq k}^{n} i \cdot y_{i} = \sum_{i=1, i \neq k}^{n} i \cdot x_i.
\end{equation}
If~\(x_{k} =0,\) then~\(CS_{er} \equiv a_1 \equiv 0 \mod {n+1}\) and if~\(x_{k} =1,\) then the discrepancy~\(CS_{er} - a_1 \equiv -k \mod{n+1} \neq 0.\) This allows us to correct the erasure.

\emph{Correcting a flip}: Suppose~\(\mathbf{y} = F_{\mathbf{g}}(\mathbf{x}) = (y_1,\ldots,y_n)\) is obtained after a single flip in the word~\(\mathbf{x}\) so that bit~\(y_k = 1-x_k\) for some~\(1 \leq k \leq n\) and~\(y_i = x_i\) for all~\(1 \leq i \neq k \leq n.\) The checksum
\begin{equation}\label{chk_flip22}
CS := \sum_{i=1}^{n} i \cdot y_{i} = \sum_{i=1, i\neq k-1} i \cdot x_{i} + k\cdot(1-x_{k}).
\end{equation}
Thus~\(CS - a_1 \equiv k(1-2x_{k})\) and if~\(x_{k} = 1,\) then~\(CS(j)-a \equiv -k \mod{n+1}\) and if~\(x_{k} = 0,\) then~\(CS(j)-a_1 \equiv k \mod{n+1}.\) This obtains the position of the flipped bit and allows us to correct the flip.

\emph{Correcting a deletion}: Suppose~\(\mathbf{y} = F_{\mathbf{g}}(\mathbf{x}) = (y_1,\ldots,y_{n-1})\) is obtained after a single deletion in the word~\(\mathbf{x}.\) The algorithm begins by computing the checksum and
weight of the received word~\(\mathbf{y}\) as
\begin{equation}\label{cs_wt}
CS(\mathbf{y}) := \sum_{i=1}^{n-1}i \cdot y_i \text{ and } w(\mathbf{y}) := \sum_{i=1}^{n-1} y_i.
\end{equation}
Suppose that bit~\(x_d\) at position~\(d\) was deleted from~\(\mathbf{x}\) to get~\(\mathbf{y};\)
i.e.,~\(\mathbf{y} = F_d(\mathbf{x}).\) The computed weight is
\begin{equation}\label{wy}
w(\mathbf{y}) = \sum_{i=1}^{d-1}x_i + \sum_{i=d+1}^{n} x_i
\end{equation}
and the corresponding checksum is
\begin{eqnarray}
CS(\mathbf{y}) &=& \sum_{i=1}^{d-1} i \cdot x_{i}  + \sum_{i=d}^{n-1}i \cdot x_{i+1} \nonumber\\
&=& \sum_{i=1}^{d-1}i \cdot x_i + \sum_{i=d}^{n-1} (i+1) \cdot x_{i+1} - \sum_{i=d}^{n-1}x_{i+1} \nonumber\\
&=& \sum_{i=1}^{n} i \cdot x_i -d \cdot x_d - \sum_{i=d+1}^{n} x_i. \nonumber
\end{eqnarray}
Thus the discrepancy in the checksum
\begin{equation}\label{dy_def}
D(\mathbf{y}) := (a-CS(\mathbf{y}))\mod {n+1} = d\cdot x_d + \sum_{i=d+1}^{n} x_i.
\end{equation}

If the deleted bit~\(x_d  =0,\) then
\[D(\mathbf{y}) = \sum_{i=d+1}^{n} x_i =\sum_{i=d}^{n-1} y_i \leq w(\mathbf{y})\]
and if the deleted bit~\(x_d  = 1,\) then
\begin{eqnarray}
D(\mathbf{y}) &=& \sum_{i=d+1}^{n} x_i  + d  \nonumber\\
&=& w(\mathbf{y}) + 1 + \sum_{i=1}^{d-1}(1-x_i) \nonumber\\
&=& w(\mathbf{y}) + 1 + \sum_{i=1}^{d-1}(1-y_i) \nonumber\\
&\geq& w(\mathbf{y})+1,
\end{eqnarray}
using~(\ref{wy}).

Therefore if~\(D(\mathbf{y}) \leq w(\mathbf{y}),\) set
\begin{equation}\label{f1_def}
f := \max\left\{j \geq 1 : \sum_{i=j}^{n-1}y_i = D(\mathbf{y})\right\}
\end{equation}
and
\begin{equation}\label{x1_def}
\mathbf{\hat{x}} := (y_1,\ldots,y_{f-1},0,y_{f}\ldots,y_{n-2},y_{n-1}).
\end{equation}
Else set
\begin{equation}\label{f2_def}
f := \min\left\{j \geq 1 : \sum_{i=1}^{j-1}(1-y_i) = D(\mathbf{y})-w(\mathbf{y})-1 \right\}
\end{equation}
and
\begin{equation}\label{x2_def}
\mathbf{\hat{x}} := (y_1,\ldots,y_{f-1},1,y_{f}\ldots,y_{n-2},y_{n-1}).
\end{equation}

\begin{algorithm}[H]
\caption{Single deletable error correction}\label{euclid2}
\SetNoFillComment

\Input{Received word $\mathbf{y}$}
\Output{Estimated word $\mathbf{\hat{x}}$}%
\CorrectErasure{$\mathbf{y}$}
{
$k \gets ErasurePosition(\mathbf{y})$\;
$D(\mathbf{y}) \gets \left(\sum_{i=1, i\neq k}^{n}i \cdot y_i - a\right) \mod(n+1)$\;
\eIf{$D(\mathbf{y}) = 0 $}
{
Output~$\mathbf{\hat{x}} = (y_1,\ldots,y_{k-1},0,y_{k+1},\ldots,y_n)$\;
}
{
Output~$\mathbf{\hat{x}} = (y_1,\ldots,y_{k-1},1,y_{k+1},\ldots,y_n)$\;
}
}
\CorrectFlip{$\mathbf{y}$}
{
$k \gets \left|\sum_{i=1}^{n}i \cdot y_i - a \right| \mod{n+1}$\;
Output~$\mathbf{\hat{x}} = (y_1,\ldots,y_{k-1},1-y_k,y_{k+1},\ldots,y_n)$\;
}
\CorrectDeletion{$\mathbf{y}$}
{
$w(\mathbf{y}) \gets \sum_{i=1}^{n-1} y_i, CS(\mathbf{y}) \gets \sum_{i=1}^{n-1}i \cdot y_i$\;
$D(\mathbf{y}) \gets (a-CS(\mathbf{y}))\mod {n+1}$\;
\eIf{$ D(\mathbf{y}) \leq w(\mathbf{y}) $}
{
Output~$\mathbf{\hat{x}}$ as in~(\ref{x1_def})\;
}
{
Output~$\mathbf{\hat{x}}$ as in~(\ref{x2_def})\;
}

}


\end{algorithm}

\subsection{Multiple deletable error correction}\label{x_mult}
We design~\(n-\)length codes capable of correcting up to~\(t_n\) deletable errors. Let~\(m\) be the largest integer such that~\(m \cdot (2t_n+1) \leq n.\)
For a bit~\(x \in \{0,1\}\) and integer~\(q \geq 1,\) let~\(x^{(q)} = (x,\ldots,x)\) be the~\(q-\)tuple with~\(x\) being repeated~\(q\) times.
If~\(\mathbf{x} = (x_1,\ldots,x_m)\) is a~\(m-\)length word, then define~\[\mathbf{x}^{(q)} = (x_1^{(q)},\ldots,x_m^{(q)})\] to be the new word obtained by repeating each bit~\(q\) times. Let~\(\mathbf{0}\) be the word consisting of~\(n- m\cdot (2t_n+1)\) zeros and define the~\(n-\)length code
\begin{equation}\label{cdel}
{\cal C}_{del} := \{(\mathbf{x}^{(2t_n+1)},\mathbf{0}): \mathbf{x} \in \{0,1\}^{m}\}.
\end{equation}

We see below that~\({\cal C}_{del}\) is capable of correcting up to~\(t_n\) deletable errors. Let~\(\mathbf{x} \in {\cal C}(t_n)\) and suppose~\(\mathbf{g} = (\mathbf{e},\mathbf{ty}) \in {\cal E}_n(k)\) is a pattern consisting of~\(k \leq t_n\) deletable errors. Further let~\(\mathbf{z} = F_{\mathbf{g}}(\mathbf{x})\) be the received word. To obtain the original word~\(\mathbf{x}\) from~\(\mathbf{z},\) we proceed as follows. First, we remove the trailing zeros from~\(\mathbf{z};\) if~\(\mathbf{z}\) contains a block of~\(l \geq 0\) consecutive zeros appearing at the end, remove~\(\min(l,n-m\cdot (2t_n+1))\) of the zeros from that block and call the new word as~\(\mathbf{y} = (y_1,\ldots,y_w).\)

Split the word~\(\mathbf{y}\) into blocks of length~\(2t_n+1\) and write
\begin{equation}\label{run_sp}
\mathbf{y} = (\mathbf{y}(1),\mathbf{y}(2),\ldots,\mathbf{y}(r)),
\end{equation}
where~\(\mathbf{y}(j), 1 \leq j \leq r-1\) has length~\(2t_n+1\) and~\(\mathbf{y}(r)\) has length at most~\(2t_n+1.\) Since there are at most~\(t_n\) deletable errors, we have that~\(r =m.\) Perform majority decision rule in each block; If~\(\mathbf{y}(j)\) has more zeros than ones, define~\(\hat{x}_j = 0\) else set~\(\hat{x}_j = 1.\) Output~\(\mathbf{\hat{x}} = (\hat{x}_1,\ldots,\hat{x}_m)\) as the estimated word.

To see that the output~\(\mathbf{\hat{x}} = \mathbf{x},\) let~\(1 \leq j \leq m\) and suppose that~\(x_j~=~1.\) Also~let~\(d(j),e(j)\) and~\(f(j)\) denote the number of deletions, erasures and flippings in blocks~\(x_i^{(2t_n+1)}, 1 \leq i \leq j,\) of the original word~\(\mathbf{x}^{(2t_n+1)}.\) Since \[d(j) + e(j) + f(j) \leq t_n,\] the block~\(\mathbf{y}(j)\) contains at least~\(t_n+1\) ones of the block~\(x_j^{(2t_n+1)} \in \mathbf{x}^{(2t_n+1)}.\)

We have summarized the algorithm in~(\ref{euclidmm}), where~\(BlockSplit(.,2t_n+1)\) splits the input into~\(r\) blocks such that~\(r-1\) blocks have size~\(2t_n+1\) and the final block has size at most~\(2t_n+1.\) Also the function~\(MajorityDecode(.)\) performs majority decoding as described above.

\begin{algorithm}[H]
\caption{Correcting multiple deletable errors}\label{euclidmm}
\SetNoFillComment

\Input{Received word $\mathbf{z}$}
\Output{Estimated word $\mathbf{\hat{x}}$}%
\MultDelErrDecode{}
{
\emph{Initialization}: $j \gets 1.$ \\
\emph{Preprocessing}: $\mathbf{y} \gets RemoveEndZeros(\mathbf{z})$, $(\mathbf{y}(1),\ldots,\mathbf{y}(r)) \gets BlockSplit(\mathbf{y},2t_n+1).$\\
\tcc{Do majority decoding}
\While{$j \leq r$}
{
$\hat{x}_j \gets MajorityDecode(\mathbf{y}(j))$ \;
$j \gets j+1$\;
}

\tcc{Get the original word}

\textbf{output} $(\hat{x}_1,\ldots,\hat{x}_r)$\;


}
\end{algorithm}

Since~\(\frac{n}{2t_n+1} -1 \leq m \leq \frac{n}{2t_n+1}\) we get that the redundancy
\begin{equation}\label{red_up_mm}
n\left(1-\frac{1}{2t_n+1}\right) \leq R({\cal C}(t_n)) = n - m \leq n\left(1-\frac{1}{2t_n+1}\right)+1.
\end{equation}


\subsection{Correcting~\(P-\)far deletable error patterns}\label{x2}
For integers~\(m \geq 2\) and~\(0 \leq a \leq m,\) we recall the VT codes~\(VT_{a}(m)\) defined in~(\ref{vt_def}).  For integers~\(P \geq 2\) and~\(n \geq P,\) write~\(n = t P + s\) where~\(0 \leq s < P-1.\) Let~\(\mathbf{0}\) and~\(\mathbf{1}\) denote the all zero and all ones words with length depending on the context. For integers~\(0 \leq a_1 \leq P\) and~\(0 \leq a_2 \leq P+s-1,\) define the code
\({\cal C}_{far} = {\cal C}_{far}(a_1,a_2,P,s,n)\) as
\begin{eqnarray}
{\cal C}_{far} &:=& \{\mathbf{x} = (\mathbf{x}(1),\ldots,\mathbf{x}(t)) : \mathbf{x}(i) \in VT_{a_1}(P) \setminus \{\mathbf{0},\mathbf{1}\}   \nonumber\\
&&\;\;\;\;\;\;\;\;\;\;\;\;\;\;\;\;\;\;\;\;\;\;\;\;\;\;\;\; \text{ for }1 \leq i \leq t-1, \mathbf{x}(t) \in VT_{a_2}(P+s)\}, \nonumber\\
\label{code_def2}
\end{eqnarray}
so that~\({\cal C}_{far}\) is obtained by appending together~\(t-1\) words from the code\\\(VT_{a_1}(P) \setminus \{\mathbf{0},\mathbf{1}\}\) and then appending a word containing~\(P+s\) bits from\\\(VT_{a_2}(P+s).\) We denote the bits in the~\(i^{th}\) word as~\(\mathbf{x}(i) = (x_1(i),\ldots,x_P(i))\) for~\(1 \leq i \leq t-1\) and let~\(\mathbf{x}(t) = (x_1(t),\ldots,x_{P+s}(t)).\) We choose optimal values of~\(a_1\) and~\(a_2\) later for minimizing the redundancy. The encoding and decoding described in this subsection hold for all values of~\(a_1\) and~\(a_2.\)

We see below that the code~\({\cal C}_{far}\) is capable of correcting all~\(3P-\)far deletable patterns. The main idea is that if~\(\mathbf{x}\) is corrupted by a~\(3P-\)far deletable error pattern, then at most one deletable error occurs in each word~\(\mathbf{x}(j)\)
in~(\ref{code_def2}). This allows us to correct the errors in a sequential manner. Let~\(\mathbf{y} = F_{\mathbf{g}}(\mathbf{x})\) where~\(\mathbf{g} = (\mathbf{e},\mathbf{ty})\) be the received word and  suppose that~\(\mathbf{x}(j), j \leq t-2\) is the first block of~\(\mathbf{x}\) to be corrupted by an error in~\(\mathbf{g};\) i.e.,~\(\mathbf{e} = (e_1,\ldots,e_n)\) is such that if~\(u = \min\{i : e_i = 1\},\) then~\((j-1)P +1 \leq u \leq jP.\) We consider the cases~\(j = t-1\) or~\(j=t\) at the end.

Divide~\(\mathbf{y} = (\mathbf{y}(1),\ldots,\mathbf{y}(t_1))\) into blocks where the~\(j^{th}\) block~\(\mathbf{y}(j) = (y_1(j),\ldots,y_P(j)), 1 \leq j \leq t_1-1\) has~\(P\) bits and~\(\mathbf{y}(t_1)\) has at least~\(P\) and at most~\(2P-1\) bits. Since the first error position~\((j-1)P +1 \leq u \leq jP\) falls in block~\(\mathbf{x}(j),\) the blocks~\(\mathbf{x}(i),1 \leq i \leq j-1\) are uncorrupted and~\(\mathbf{y}(i) = \mathbf{x}(i),1 \leq i \leq j-1.\) Consequently, the checksum
\begin{equation}\label{check_sum_i}
CS(l) := \sum_{i=1}^{P}i \cdot y_i(l) \equiv a_1 \mod(P+1)
\end{equation}
for~\(1 \leq l \leq j-1\) and so the checksum difference~\(|CS(l) - a_1| \mod(P+1) \equiv 0\) for all blocks~\(1 \leq l \leq j-1.\)

If the error corrupting the block~\(\mathbf{x}(j)\) is an erasure, then~\(y_u(j) = \varepsilon\) (the erasure symbol) and~\(y_k(j) = x_k(j)\) for all~\(1 \leq k \neq u \leq P.\) Therefore the modified checksum
\begin{equation}\label{er_chk_sum}
CS_{er}(j) := \sum_{i=1, i \neq u}^{P} i \cdot y_{i}(j) = \sum_{i=1, i \neq u}^{P} i \cdot x_i(j).
\end{equation}
If~\(x_{u}(j) =0,\) then~\(CS_{er}(j) - a_1 \equiv 0 \mod (P+1)\) and if~\(x_{u}(j) =1,\) then~\(CS_{er}(j) - a_1 \equiv -u \mod(P+1) \neq 0\) and so the modified checksum difference \[|CS_{er}(j) - a_1| \mod(P+1) \equiv u \ind(x_u(j) = 1)\] is nonzero if and only if the erased bit is a one. This allows us to correct the erasure. 

If the error corrupting the block~\(\mathbf{x}(j)\) is a flip or deletion, then it is not directly detectable as in the case of erasure above. We therefore use the checksums of both~\(\mathbf{y}(j)\) and~\(\mathbf{y}(j+1)\) to indirectly deduce the nature of the error. Indeed, if the error is a flip then~\(y_u(j) = 1-x_{u}(j)\) and~\(y_k(j) = x_k(j)\) for~\(1 \leq k \neq u \leq P\) and so the checksum of the block~\(\mathbf{y}(j)\) is
\begin{equation}\label{chk_flip}
CS(j) := \sum_{i=1}^{P} i \cdot y_{i}(j) = \sum_{i=1, i\neq u} i \cdot x_{i}(j) + u\cdot(1-x_{u}(j)).
\end{equation}
Thus~\(CS(j) - a_1 \equiv u(1-2x_{u}(j))\) and so the checksum difference for the block~\(\mathbf{y}(j),\) \[|CS(j)-a_1| \equiv u \mod(P+1)\]  is nonzero and equals the location of the flipped bit. Moreover, since the errors are~\(3P-\)far apart, no errors have corrupted the block~\(\mathbf{x}(j+1)\) of the original word (which also contains~\(P\) bits since~\(j \leq t-2\)). Therefore~\(\mathbf{y}(j+1) = \mathbf{x}(j+1)\) and the checksum difference~\(|CS(j+1)-a_1| \mod(P+1)\) for the block~\(\mathbf{y}(j+1)\) is zero.

If on the other hand the error corrupting the block~\(\mathbf{x}(j)\) is a deletion, we may or may not get a nonzero checksum difference in block~\(\mathbf{y}(j).\) But we are guaranteed a nonzero checksum difference in block~\(\mathbf{y}(j+1).\) This is because in the block~\(\mathbf{y}(j+1),\) the corresponding bits of~\(\mathbf{x}(j+1)\) are shifted one position to the left; i.e.,~\(y_i(j) = x_{i+1}(j)\) for~\(1 \leq i \leq P-1.\) Therefore the checksum
\begin{eqnarray}
CS(j+1) &=& \sum_{i=1}^{P-1}i\cdot x_{i+1}(j+1) + P\cdot y_P(j+1) \nonumber\\
&=& \sum_{i=1}^{P-1} (i+1) \cdot x_{i+1}(j+1) + P\cdot y_P(j+1) - \sum_{i=2}^{P} x_i(j+1) \nonumber\\
&=& \sum_{i=1}^{P}i \cdot x_i(j+1) + P\cdot y_P(j+1) - \sum_{i=1}^{P} x_i(j+1) \nonumber
\end{eqnarray}
and the checksum difference~\[CS(j+1) - a_1 \equiv P\cdot y_P(j+1) - \sum_{i=1}^{P} x_i(j+1) \mod(P+1).\] If~\(y_P(j+1) = 1,\) then
\[P \geq P - \sum_{i=1}^{P} x_i(j+1) \geq 1,\] since not all bits in a block can be one. If~\(y_P(j+1) = 0,\)
then \[-P \leq - \sum_{i=1}^{P} x_i(j+1) \leq -1,\] since not all bits in a block can be zero. In other words, the checksum difference for the block~\(\mathbf{y}(j+1),\)~\(|CS(j+1)-a_1| \mod(P+1) \neq 0.\)





Summarizing, suppose we get the first nonzero checksum difference (mismatch) at some block~\(\mathbf{y}(j)\) containing~\(P\) bits. The mismatch may be because of an error either in block~\(\mathbf{y}(j-1)\) or block~\(\mathbf{y}(j).\) If indeed the mismatch was because of an error in block~\(\mathbf{y}(j-1),\) then the error must be a deletion since a flipping would have already caused a mismatch for block~\(\mathbf{y}(j-1)\) (see discussion following~(\ref{chk_flip})). We therefore remove the last bit of the block~\(j-1\) and perform single deletion correction to get a new word~\(\mathbf{\hat{x}}(j-1).\) If~\(\mathbf{\hat{x}}(j-1) \neq \mathbf{y}(j-1),\) we replace the first~\(P-1\) bits of block~\(\mathbf{y}(j-1)\) with the block~\(\mathbf{\hat{x}}(j-1).\) 

If~\(\mathbf{\hat{x}}(j-1) = \mathbf{y}(j-1),\) then error necessarily has occurred in block~\(\mathbf{y}(j)\) and we need to use block~\(\mathbf{y}(j+1)\) to determine the nature of error as described above. If~\(\mathbf{y}(j+1)\) is not the final block in~\(\mathbf{y},\) then we proceed as above and deduce that the error in~\(\mathbf{y}(j)\) is either a flipping or deletion depending on whether the checksum difference of the block~\(\mathbf{y}(j+1)\) is zero or not, respectively. 

If on the other hand~\(\mathbf{y}(j+1)\) is the final block in~\(\mathbf{y},\) then it is necessarily true that~\(\mathbf{y}\) is obtained after corrupting~\(\mathbf{x}\) by a single error occurring in one of the last two blocks~\(\mathbf{x}(t-1)\) or~\(\mathbf{x}(t);\) i.e., the first error position~\(u\) defined in the paragraph prior to~(\ref{check_sum_i}) falls in one of the last two blocks~\(\mathbf{x}(t-1)\) or~\(\mathbf{x}(t).\) Since errors are~\(3P-\)far apart, the error at position~\(u\) is the only error corrupting~\(\mathbf{x}\) and so the corrupted word~\(\mathbf{y}\) has either~\(n\) or~\(n-1\) bits.

Split~\(\mathbf{y} = (\mathbf{y}(1),\ldots,\mathbf{y}(t-1),\mathbf{y}(t))\) where the first~\(t-1\) blocks contain~\(P\) bits each and the final block~\(\mathbf{y}(t)\) has either~\(P+s\) or~\(P+s-1\) bits. One of the blocks in~\(\{\mathbf{y}(t-1),\mathbf{y}(t)\}\) has been corrupted by an error
and we perform correction as follows. If the error is an erasure, then we correct the erased bit simply by computing the checksum difference as described before (see discussion prior to~(\ref{er_chk_sum})). If there is no erasure but the received word~\(\mathbf{y}\) has~\(n\) bits, then the error is a flipping and we obtain the location of the flipping by computing the checksum difference for each of the blocks~\(\mathbf{y}(t-1)\) and~\(\mathbf{y}(t),\) as described in the paragraph containing~(\ref{chk_flip}).


Finally, if~\(\mathbf{y}\) has~\(n-1\) bits, then the error is a deletion and we determine the block where the deletion has occurred as follows. We perform single deletion correction on the first~\(P-1\) bits~\(\{\mathbf{y}_i(t-1)\}_{1 \leq i \leq P-1}\) of~\(\mathbf{y}(t-1)\) and get a new~\(P\) bit word~\(\mathbf{\hat{w}}.\) If~\(\mathbf{\hat{w}} \neq \mathbf{y}(t-1),\) then the deletion necessarily has occurred in the block~\(\mathbf{x}(t-1)\) of the original word~\(\mathbf{x}.\) We therefore finish the iteration by inserting~\(\mathbf{\hat{w}}\) into the word~\(\mathbf{y}\) giving the output \[\mathbf{\hat{x}} = (\mathbf{y}(1),\ldots,\mathbf{y}(t-2),\mathbf{\hat{w}},y_P(t-1),\mathbf{y}(t)).\]

If on the other hand~\(\mathbf{\hat{w}} = \mathbf{y}(t-1),\) then the deletion has occurred
in the final block~\(\mathbf{x}(t)\) of the original word~\(\mathbf{x}.\)  We then perform single deletion correction on the~\(P+s-1\) bits of~\(\mathbf{y}(t)\) to get a new word~\(\mathbf{\hat{z}}\) containing~\(P+s\) bits and output \[\mathbf{\hat{x}} = (\mathbf{y}(1),\ldots,\mathbf{y}(t-1),\mathbf{\hat{z}}).\]

The procedure described above corrects a single error in the corrupted word~\(\mathbf{y}\) and after the iteration, the first error occurring at position~\(u\) in the error pattern~\(\mathbf{g}\) (see discussion prior to~(\ref{check_sum_i})) is corrected. The received word after the first iteration~\(\mathbf{y}^{(1)} = F_{\mathbf{g}(1)}(\mathbf{x})\) is a lesser corrupted version of~\(\mathbf{x}\) than~\(\mathbf{y}\) and~\(\mathbf{g}(1) = (\mathbf{e}(1),\mathbf{ty})\) is an error pattern whose first error occurs strictly after the first error in~\(\mathbf{g};\) i.e., if~\(\mathbf{e}(1) = (e_1(1),\ldots,e_n(1))\) and~\(u_1 = \min\{i:e_i(1) = 1\},\) then~\(u_1 > u.\)

We now repeat the above procedure to correct the error at position~\(u_1\) and continuing iteratively, we sequentially correct all the remaining errors. The procedure stops if there is no checksum mismatch and we output the final word as our estimate of~\(\mathbf{x}.\) We summarize the method in Algorithm~\ref{euclid} below, where the function~\(BlockSplit(.,P)\) splits the input into blocks of length~\(P\) with the last block containing between~\(P\) and~\(2P-1\) bits.



Finally, to compute the redundancy of the code~\({\cal C}_{far},\) we choose~\(a_1\) and~\(a_2\) such that
\begin{equation}\label{a1_a2_choose}
\#VT_{a_1}(P) \geq \frac{2^{P}}{P+1} \text{ and } \#VT_{a_2}(s) \geq \frac{2^{P+s}}{P+s+1}.
\end{equation}
Since we append together~\(t-1\) words each chosen from~\(VT_{a_1}(P) \setminus \{\mathbf{0},\mathbf{1}\}\) and one word from~\(VT_{a_2}(P+s),\) we have that
\begin{eqnarray}
\#{\cal C}_{far} &\geq& \left(\frac{2^{P}}{P+1}-2\right)^{t-1} \frac{2^{P+s}}{(P+s+1)} \nonumber\\
&=& \left(\frac{2^{P}}{P+1}\right)^{t-1}(1-\delta(P))^{t-1} \left(\frac{2^{P+s}}{P+s+1}\right) \nonumber\\
&=& 2^{n} \left(\frac{1-\delta(P)}{P+1}\right)^{t-1} \frac{1}{P+s+1}, \label{cfar_1}
\end{eqnarray}
where~\(\delta(P) = \frac{P+1}{2^{P-1}}\) is as in~(\ref{del_def_main}) and relation~(\ref{cfar_1}) is true  since~\(Pt + s = n.\)
Using~\(s+1 \leq P,\) we further get
\[\#{\cal C}_{far} \geq 2^{n} \left(\frac{1-\delta(P)}{P+1}\right)^{t-1} \frac{1}{2P} \geq 2^{n} \left(\frac{1-\delta(P)}{P+1}\right)^{\frac{n}{P}-1} \frac{1}{2P},\]
since~\(t \leq \frac{n}{P}\) and~\(\frac{1-\delta(P)}{P+1} < 1.\) Thus the redundancy
\begin{equation}\label{rc2}
R({\cal C}_{far}) \leq \left(\frac{n}{P}-1\right)\log\left(\frac{P+1}{1-\delta(P)}\right) + \log{P} + 1
\end{equation}
for all~\(n\geq P \geq 2.\)~\(\qed\)


\begin{algorithm}[tbp]\footnotesize
\caption{Correcting a~\(3P-\)far deletable error pattern}\label{euclid}
\SetNoFillComment

\Input{Received word $\mathbf{y}$}
\Output{Estimated word $\mathbf{\hat{x}}$}%
\SequentialDecode{}
{
\emph{Initialization}: $\mathbf{\hat{x}} \gets \mathbf{y}, j \gets 0, CS  \gets 0.$ \\
\emph{Preprocessing}: $(\mathbf{\hat{x}}(1),\ldots,\mathbf{\hat{x}}(t-1),\mathbf{\hat{x}}(t)) \gets BlockSplit(\mathbf{\hat{x}},P),$\\
\emph{where $P \leq Length(\mathbf{\hat{x}}(t)) < 2P$ }.\\
\tcc{Pass all blocks with checksum match}
\While {$CS = 0$ and $j \leq t$ }{
    $j \gets j+1, CS \gets -1$\;
    \If{erasure in~$j^{th}$ block}
    {$\mathbf{\hat{x}}(j) \gets CorrectErasure(\mathbf{\hat{x}}(j))$\;}
    \If{$j < t$ or $j=t$ and $Length(\hat{\mathbf{x}}(t)) = P+s$}
        {$CS \gets CheckSumDifference(\mathbf{\hat{x}}(j))$\;}

}


\If{$CS = 0$ and $j=t$ }
{
\textbf{output} $(\mathbf{\hat{x}}(1),\ldots,\mathbf{\hat{x}}(t-1),\mathbf{\hat{x}}(t))$\;
}

\tcc{Checksum mismatch in~\(j^{th}\) block; Check if deletion in~\((j-1)^{th}\) block}
$l \gets \min(j-1,1),\mathbf{T} \gets CorrectDeletion(RemoveLastBit(\mathbf{\hat{x}}(l)))$\; 
\eIf{$\mathbf{T} \neq \mathbf{\hat{x}}(l) $}  
{$\mathbf{T} \gets (\mathbf{T},LastBit(\mathbf{\hat{x}}(l)))$\;}
{
\tcc{Error in~\(j^{th}\) block; If final block then directly perform correction}
\If{$j = t$}
{
\eIf{$\mathbf{\hat{x}}(j)$ has $P+s$ bits}
{$\mathbf{T} \gets CorrectFlip(\mathbf{\hat{x}}(j))$\;}
{$\mathbf{T} \gets CorrectDeletion(\mathbf{\hat{x}}(j))$\;}
{\textbf{output} $(\mathbf{\hat{x}}(1),\ldots,\mathbf{\hat{x}}(t-1),\mathbf{T})$\;}
}

\tcc{In all other cases, determine if error is flipping or deletion using block~\(j+1\)}
 $l \gets j, ErrorType \gets 1$\;
 \If{$l <t-1 $ or $l = t-1$ and $Length(\mathbf{\hat{x}}(l+1)) = P+s$}
 {$ErrorType \gets CheckSumDifference(\mathbf{\hat{x}}(l+1))$\;}
 \eIf{$ErrorType = 0$}
 {$\mathbf{T} \gets CorrectFlip(\mathbf{\hat{x}}(l))$\;}
 {$\mathbf{T} \gets CorrectDeletion(RemoveLastBit(\mathbf{\hat{x}}(l)))$\;
 $\mathbf{T} \gets (\mathbf{T},LastBit(\mathbf{\hat{x}}(l)))$\;
 }
}

\tcc{Update the estimate~\(\mathbf{\hat{x}}\) with the new block $\mathbf{T}$}
{
$\mathbf{\hat{x}} \gets (\mathbf{\hat{x}}(1),\ldots,\mathbf{\hat{x}}(l-1),\mathbf{T},\mathbf{\hat{x}}(l+1),\ldots,\mathbf{\hat{x}}(t))$\;
\textbf{goto} \emph{Preprocessing}\;
}


}
\end{algorithm}

\setcounter{equation}{0}
\renewcommand\theequation{\thesection.\arabic{equation}}
\section{Proof of Theorems}\label{pf2}
We prove the lower bound on the redundancy for all the Theorems in Section~\ref{pf_red_min}. We prove the rest in this section.




\emph{Proof of Theorem~\ref{prop1}}: From Algorithm~\ref{euclidmm}, the code~\({\cal C}_{del}\) as described in~(\ref{cdel}) is capable of correcting up to~\(t_n\) deletable errors and the upper bound on the redundancy is obtained from~(\ref{red_up_mm}).~\(\qed\)

\emph{Proof of Theorem~\ref{thm_11}}: For~\(n \geq 3,\) we write~\(n = tP + s\) where~\(0 \leq s < P\) and set~\({\cal C}_{far} = {\cal C}_{far}(a_1,a_2,P,s)\) to be the code defined in~(\ref{code_def2}). From Algorithm~\ref{euclid} we have that~\({\cal C}_{far}\) is capable of correcting all~\(3P-\)far deletable error pattern and from~(\ref{rc2}) we also obtain the upper bound~(\ref{red_upp44}) on the redundancy.~\(\qed\)


\emph{Proof of Theorem~\ref{thm3}}: From Algorithm~\ref{euclidmm}, the code~\({\cal C}_{bur} := {\cal C}_{del}\) as described in~(\ref{cdel}) is capable of
correcting up to~\(b\) deletable errors and the upper bound on the redundancy is obtained from~(\ref{red_up_mm}), with~\(t_n = b.\)~\(\qed\)

\emph{Proof of Theorem~\ref{thm_12}}: For~\(n \geq 3\) define
\begin{equation}\label{pn_def}
P_n := \frac{n}{t_n^2 \omega_n}
\end{equation}
and assume without loss of generality that~\(P_n\) is an integer for all~\(n.\)\\Using~\(\frac{\omega_n t_n^3}{n} \longrightarrow 0\) (see~(\ref{omega_cond})), we also get that~\(P_n \longrightarrow \infty\) as~\(n \rightarrow \infty.\) Let~\({\cal C}_{frac} := {\cal C}_{far}(n,P_n)\) be the code defined in Theorem~\ref{thm_11} so that~\({\cal C}_{frac}\) is capable of correcting all~\(3P_n-\)far deletable error patterns. From~(\ref{red_upp44}), we also get that the redundancy of~\({\cal C}_{frac}\)  is at most
\begin{eqnarray}
&&\left(\frac{n}{P_n}-1\right) \log\left(\frac{P_n+1}{1-\delta(P_n)}\right) + \log{P_n} + 1 \nonumber\\
&&=\frac{n}{P_n} \log{P_n} + \left(\frac{n}{P_n}-1\right)\log\left(\frac{1+\frac{1}{P_n}}{1-\delta(P_n)}\right) + 1. \nonumber
\end{eqnarray}
Using~\(P_n \longrightarrow \infty\) (see statement following~(\ref{pn_def})) and the fact that\\\(\max(|\log(1+x)|,|\log(1-x)|) \leq 2x\) for all~\(x\) small,
\begin{equation}\nonumber
R({\cal C}_{frac}) \leq \frac{n}{P_n}\log{P_n} + \left(\frac{n}{P_n}-1\right)\left(\frac{2}{P_n} + 2\delta(P_n)\right) \leq \frac{n}{P_n}\log{P_n} + \frac{n}{P_n},
\end{equation}
for all~\(n\) large, since~\(P_n \longrightarrow \infty\) and so~\(\frac{2}{P_n}+2\delta(P_n) \longrightarrow 0\) as~\(n \rightarrow \infty.\) Thus
\begin{equation}\label{rqn1}
R({\cal C}_{frac}) \leq \frac{n}{P_n}\log{P_n} + \frac{n}{P_n} = \frac{n}{P_n} \log(2P_n) = \omega_n t_n^2 \log\left(\frac{2n}{\omega_n t_n^2}\right),
\end{equation}
proving property~\((a1)\) of Theorem~\ref{thm_12}.

We show below that if~\({\cal F}_n\) is the set of all~\(3P_n-\)far error patterns, then
\begin{equation}\label{frac_pattern}
\frac{\#{\cal F}_n}{\#{\cal E}_n(t_n)} \geq 1-\frac{42}{\omega_n}
\end{equation}
for all~\(n\) large. Since~\({\cal Q}_n\) is capable of correcting all~\(3P_n-\)far deletable error patterns, this proves property~\((a2)\) of Theorem~\ref{thm_12}.

First, the total number of error patterns in~\({\cal E}_n(t_n)\) is
\begin{equation}\label{en_tot}
\#{\cal E}_n(t_n)  = \sum_{k=0}^{t_n} {n \choose k} 3^{k}.
\end{equation}
Suppose~\(\mathbf{g} = (\mathbf{e},\mathbf{ty}),\mathbf{e} = (e_1,\ldots,e_n)\) is not a~\(3P_n-\)far deletable error pattern. Let~\(n = 3LP_n + s, 0 \leq s \leq 3P_n-1\) and split~\(\mathbf{e} = (\mathbf{e}(1),\ldots,\mathbf{e}(L))\) into~\(L\) blocks, where the first~\(L-1\) blocks each have length~\(3P_n\) and the final block has length~\(3P_n+s.\) Since~\(\mathbf{g}\) is not a~\(3P_n-\)far error pattern, there are only two possibilities:\\
\((p1)\) One of the~\(L\) blocks in~\(\mathbf{e}\) contains at least two nonzero entries.\\
\((p2)\) Two consecutive blocks in~\(\mathbf{e}\) contain exactly one nonzero entry each.\\
We compute the number of patterns for each possibility.

To compute the number of patterns~\(\#{\cal F}(p1)\) in possibility~\((p1),\) we argue as follows: If~\({\cal S}(l), 1 \leq l \leq L\) is the set of error patterns~\(\mathbf{g} =(\mathbf{e},\mathbf{ty})\) where the~\(l^{th}\) block~\(\mathbf{e}(l)\) of~\(\mathbf{e}\) has two or more nonzero entries, then
\begin{equation}\label{fp1_12}
\#{\cal F}(p1) \leq \sum_{l=1}^{L}\#{\cal S}(l) = (L-1) \#{\cal S}(1) + \#{\cal S}(L).
\end{equation}

We estimate~\(\#{\cal S}(1)\) as follows. Suppose~\(\mathbf{g} = (\mathbf{e},\mathbf{ty}) \in {\cal S}(1)\) and suppose that~\(\mathbf{e} = (e_1,\ldots,e_n)\) has~\(k \leq t_n\) nonzero entries. We split~\(\mathbf{e} = (\mathbf{e}(1),\ldots,\mathbf{e}(L))\) into~\(L\) blocks as before. Among the~\(3P_n\) bits in the first block~\(\mathbf{e}(1),\) we choose~\(2 \leq r \leq k\) of the bits in~\({3P_n \choose r}\) ways and choose~\(k-r\) bits of the remaining~\(n-3P_n\) bits in~\({n-3P_n \choose k-r}\) ways. The binomial coefficient~\({3P_n \choose r}\)  is nonzero since~\(r \leq k \leq t_n\) and~\(P_n = \frac{n}{t_n^2 \omega_n} \geq t_n\) for all~\(n\) large, since~\(\frac{\omega_n t_n^3}{n} \longrightarrow 0\) (see~(\ref{omega_cond})). Thus
\begin{equation}\nonumber
\#{\cal S}(1) = \sum_{k=2}^{t_n} \sum_{r=2}^{k} {n-3P_n \choose k-r} {3P_n \choose r} 3^{k}  \nonumber
\end{equation}
and using~\(\sum_{k=0}^{r} {x \choose k-r}{y \choose r} = {x+y \choose k}\) with~\(x = n-3P_n\) and~\(y = 3P_n,\) we get
\begin{equation}\label{fp11}
\#{\cal S}(1) = \sum_{k=2}^{t_n} \left(1-S(n,k)\Delta\right) {n \choose k} 3^{k} , \nonumber
\end{equation}
where~\(S(n,k) = \frac{{n-3P_n \choose k}}{{n \choose k}}\) and
\begin{equation}
\Delta = 1 + \frac{{3P_n \choose 1} \cdot {n-3P_n \choose k-1}}{{n-3P_n \choose k}} = 1 + \frac{3P_nk}{n-3P_n-k+1} \geq 1 + \frac{3P_n k}{n}. \label{del1del2}
\end{equation}

We now evaluate~\(S(n,k)\) by letting~\(x = n-3P_n\) to get
\begin{equation}\label{snk_def2}
S(n,k) = \frac{{x \choose k}}{{n \choose k}}=  \frac{x(x-1)\ldots(x-k+1)}{n(n-1)\ldots(n-k+1)} = \left(\frac{x}{n}\right)^{k}\frac{L(x,k)}{L(n,k)},
\end{equation}
where
\begin{equation}\label{lx_def}
1 \geq L(x,k) := \prod_{i=0}^{k-1} \left(1-\frac{i}{x}\right) \geq \left(1- \frac{1}{x}\sum_{i=1}^{k-1}i\right) = 1-\frac{k(k-1)}{2x}.
\end{equation}
The term~\(x = n-3P_n  = n -\frac{3n}{t_n^2 \omega_n} \geq \frac{n}{2}\) since~\(\omega_n \geq 6.\) Therefore
\[1-\frac{k(k-1)}{2x} = 1- \frac{k(k-1)}{2n} - \frac{3P_nk(k-1)}{2nx} \geq 1- \frac{k(k-1)}{2n} - \frac{3P_nk^2}{n^2}\]
and so
\begin{equation}\label{lx_est1}
L(x,k) \geq 1- \frac{k(k-1)}{2n} - \frac{3P_nk^2}{n^2}.
\end{equation}

To find an upper bound for~\(L(n,k),\) we use the fact that~\(1-y <e^{-y}\) for~\(0 < y< 1\) to get
\[\log{L(n,k)} \leq -\sum_{i=1}^{k-1} \frac{i}{n}  = -\frac{k(k-1)}{2n}.\] But~\(\frac{k(k-1)}{2n} \leq \frac{t_n^2}{2n}\longrightarrow 0\) since~\(\frac{\omega_nt_n^3}{n} \longrightarrow 0\) as~\(n \rightarrow \infty\) by~(\ref{omega_cond}) and so we get
\begin{eqnarray}
L(n,k) &\leq& \exp\left(-\frac{k(k-1)}{2n}\right) \nonumber\\
&\leq& 1- \frac{k(k-1)}{2n} + \left(\frac{k(k-1)}{2n}\right)^2  \nonumber\\
&\leq& 1- \frac{k(k-1)}{2n} + \frac{k^4}{n^2} \label{lnk_up}
\end{eqnarray}
for all~\(n \geq N\) large, where~\(N\) does not depend on~\(k.\) Using~(\ref{lnk_up}) and the lower bound~(\ref{lx_est1}) we have
\begin{equation} \label{l_rat}
\frac{L(x,k)}{L(n,k)} -1 \geq \frac{1-\frac{k(k-1)}{2n} - \frac{3P_nk^2}{n^2}}{1- \frac{k(k-1)}{2n} + \frac{k^4}{n^2}} - 1  = -\frac{3P_nk^2 + k^4}{n^2}\left(1- \frac{k(k-1)}{2n} + \frac{k^4}{n^2}\right)^{-1}
\end{equation}
Since~\(\frac{t_n^2}{n} \leq \frac{\omega_n t_n^3}{n} \longrightarrow 0\) as~\(n \rightarrow \infty\) by~(\ref{omega_cond}) and~\(k \leq t_n,\) we have \[1- \frac{k(k-1)}{2n} + \frac{k^4}{n^2} \geq 1-\frac{t_n^2}{2n} \geq \frac{1}{2}\] for all~\(n\) large and so we get from~(\ref{l_rat}) that
\begin{equation} \label{l_rat2}
\frac{L(x,k)}{L(n,k)} -1 \geq -2\frac{3P_nk^2 + k^4}{n^2}.
\end{equation}

Substituting~(\ref{l_rat2}) into~(\ref{snk_def2}) and using the fact that~\((1-x)^{r} \geq 1-rx\) for~\(r \geq 1\) and~\(0 < x< 1,\) we get that
\begin{eqnarray}
S(n,k) &\geq& \left(1-\frac{3P_n}{n}\right)^{k}\left(1-2\frac{3P_nk^2 + k^4}{n^2}\right) \nonumber\\
&\geq& \left(1-\frac{3P_nk}{n}\right)\left(1-2\frac{3P_nk^2 + k^4}{n^2}\right) \label{phjj_111_w}\\
&\geq& 1-\frac{3P_n k}{n} - 2\frac{3P_nk^2 + k^4}{n^2}. \label{phjj_111}
\end{eqnarray}
Using~(\ref{phjj_111}) and~(\ref{del1del2}), we therefore get
\begin{eqnarray}
S(n,k)\Delta &\geq& \left(1-\frac{3P_n k}{n} - 2\frac{3P_nk^2 + k^4}{n^2}\right)\left(1 + \frac{3P_n k}{n}\right) \nonumber\\
&=&1-\left(\frac{3P_n k}{n}\right)^2 -2\frac{3P_nk^2 + k^4}{n^2}\left(1 + \frac{3P_n k}{n}\right). \nonumber
\end{eqnarray}
Since~\(k \leq t_n\) we have that~\(\frac{P_n k}{n} \leq \frac{P_n t_n}{n} = \frac{1}{t_n \omega_n} \leq 1,\) because~\(\omega_n\) and~\(t_n\) are both at least one. Thus~\(1+\frac{3P_n k}{n} \leq 4\)
and so
\begin{equation}\label{snk_low1}
S(n,k)\Delta \geq 1-\frac{9P_n^2k^2}{n^2} - 8 \frac{3P_nk^2 + k^4}{n^2}.
\end{equation}

Substituting~(\ref{snk_low1}) into~(\ref{fp11}), we get
\begin{eqnarray}
\#{\cal S}(1) &\leq& \sum_{k=2}^{t_n}\left(\frac{9P_n^2k^2}{n^2} + 8 \frac{3P_nk^2 + k^4}{n^2}\right) {n \choose k} 3^{k} \nonumber\\
&\leq& \left(\frac{9P_n^2t_n^2}{n^2} + 8 \frac{3P_nt_n^2 + t_n^4}{n^2}\right) \sum_{k=2}^{t_n}{n \choose k} 3^{k} \nonumber\\
&\leq& \left(\frac{9P_n^2t_n^2}{n^2} + 8 \frac{3P_nt_n^2 + t_n^4}{n^2}\right) (\#{\cal E}_n(t_n)),\nonumber
\end{eqnarray}
by~(\ref{en_tot}). The above estimate holds for all the sets~\({\cal S}(l), 1 \leq l \leq L-1\) (see discussion prior to~(\ref{fp1_12})). Performing the same analysis with~\(3P_n+s\) instead of~\(3P_n\) and using the fact that~\(s \leq 3P_n,\)
we get \[\#{\cal S}(L) \leq \left(\frac{(3P_n+s)^2t_n^2}{n^2} + 8\frac{(3P_n+s)t_n^2 + t_n^4}{n^2}\right) (\#{\cal E}_n(t_n)).\]
Thus
\[\#{\cal S}(l) \leq \left(\frac{(6P_n)^2t_n^2}{n^2} + 8 \frac{6P_nt_n^2 + t_n^4}{n^2}\right) (\#{\cal E}_n(t_n)) \leq \frac{84 P_n^2 t_n^2 + 8t_n^4}{n^2} (\#{\cal E}_n(t_n)) \]
for all~\(1 \leq l \leq L\) and since~\(L \leq \frac{n}{3P_n},\)
we get from~(\ref{fp1_12}) that
\begin{equation}\label{fp1_122}
\#{\cal F}(p1) \leq \frac{n}{3P_n}\left(\frac{84P_n^2t_n^2 + 8t_n^4}{n^2}\right) (\#{\cal E}_n(t_n)) = \left(\frac{28P_nt_n^2}{n} + \frac{8t_n^4}{3nP_n}\right) (\#{\cal E}_n(t_n)).
\end{equation}
Since~\(P_n = \frac{n}{t_n^2 \omega_n}\) and~\(\frac{\omega_n t_n^3}{n} \longrightarrow 0\) as~\(n \rightarrow \infty\) (see~(\ref{omega_cond})), we get that
\[\frac{28P_nt_n^2}{n} + \frac{4t_n^4}{3nP_n}  = \frac{28}{\omega_n} + \frac{4}{3\omega_n} \left(\frac{\omega_n t_n^3}{n}\right)^2 \leq \frac{30}{\omega_n},\]
for all~\(n\) large.


To compute the number of patterns~\(\#{\cal F}(p2)\) in possibility~\((p2)\) described prior to~(\ref{fp1_12}), we argue as follows: If~\({\cal T}(l)\) is the set of error patterns~\(\mathbf{g} =(\mathbf{e},\mathbf{ty})\) where the blocks~\(\mathbf{e}(l)\) and~\(\mathbf{e}(l+1),1 \leq l \leq L-1\) each have exactly one nonzero entry each, then
\begin{equation}\label{fp2}
\#{\cal F}(p2) \leq \sum_{l=1}^{L-1}\#{\cal T}(l) = (L-2) \#{\cal T}(1) + \#{\cal T}(L-1).
\end{equation}

We estimate~\(\#{\cal T}(1)\) as follows. Suppose~\(\mathbf{g} = (\mathbf{e},\mathbf{ty}) \in {\cal T}(1)\) and suppose that~\(\mathbf{e} = (e_1,\ldots,e_n)\) has~\(k\) nonzero entries. We split~\(\mathbf{e} = (\mathbf{e}(1),\ldots,\mathbf{e}(L))\) into~\(L\) blocks as before. Among the~\(3P_n\) bits in the first block~\(\mathbf{e}(1),\) we choose one bit, among the~\(3P_n\) bits of~\(\mathbf{e}(2)\) we choose one bit and choose~\(k-2\) bits from the remaining~\(n-6P_n\) bits of~\(\mathbf{e}(j), j \neq 1,2.\) This can be done in~\({3P_n \choose 1}{3P_n \choose 1}{n-6P_n \choose k-2}\) ways and so
\begin{equation}
\#{\cal T}(1) = \sum_{k=2}^{t_n}  9P_n^2{n-6P_n \choose k-2} 3^{k} \leq 9P_n^2 \sum_{k=2}^{t_n} {n \choose k-2} 3^{k} . \label{t1_est1}
\end{equation}
For~\(k \leq t_n\) the ratio
\begin{equation}\label{nk_est1}
\frac{{n \choose k-2}}{{n \choose k}} = \frac{k(k-1)}{(n-k+2)(n-k+1)} \leq \frac{t_n^2}{(n-t_n+2)^2}=\left(\frac{t_n}{n}\right)^2 \left(1-\frac{t_n-2}{n}\right)^{-2}
\end{equation}
and using~\(\frac{t_n}{n} \leq \frac{\omega_n t_n^3}{n} \longrightarrow 0\) as~\(n \rightarrow \infty\) (see~(\ref{omega_cond})), we also have that\\\(\left(1-\frac{t_n-2}{n}\right)^{-2} \leq 2\) for all~\(n\) large.
From~(\ref{t1_est1}), we therefore get that
\begin{equation}
\#{\cal T}(1) \leq \frac{18P_n^2 t_n^2}{n^2}\sum_{k=2}^{t_n} {n \choose k} 3^{k} \leq \frac{18P_n^2 t_n^2}{n^2} (\#{\cal E}_n(t_n)) \label{t1_est12}
\end{equation}
using~(\ref{en_tot}).

Performing an analogous analysis as above for the last two blocks with length~\(3P_n\) and~\(3P_n+s,\) we get
\begin{equation}
\#{\cal T}(L-1) = \sum_{k=2}^{t_n}  3P_n(3P_n+s) {n-6P_n-s \choose k-2} 3^{k} \leq 18P_n^2 \sum_{k=2}^{t_n} {n \choose k-2} 3^{k}, \label{t2_est}
\end{equation}
since~\(s \leq 3P_n.\) Again using~(\ref{nk_est1}), we get
\begin{equation}
\#{\cal T}(L-1) \leq \frac{36P_n^2 t_n^2}{n^2}\sum_{k=2}^{t_n} {n \choose k} 3^{k} \leq \frac{36P_n^2 t_n^2}{n^2} (\#{\cal E}_n(t_n)) \label{t1_est121}
\end{equation}
using~(\ref{en_tot}). Using~(\ref{t1_est12}) and~(\ref{t1_est121}) in~(\ref{fp2}) and we therefore get that
\begin{equation}\label{fp2_est}
\#{\cal F}(p2) \leq L \cdot \frac{36P_n^2 t_n^2}{n^2} (\#{\cal E}_n(t_n))  \leq \frac{12P_n t_n^2}{n} (\#{\cal E}_n(t_n)) =\frac{12}{\omega_n}  (\#{\cal E}_n(t_n)),
\end{equation}
where the second inequality in~(\ref{fp2_est}) is true since~\(L \leq \frac{n}{3P_n}\)  and the final estimate in~(\ref{fp2_est}) is true since~\(P_n =\frac{n}{t_n^2 \omega_n}\) (see~(\ref{pn_def})).

Recall from discussion prior to~(\ref{fp1_12}) that~\({\cal F}_n\) denotes the set of all~\(3P_n-\)far error patterns and using~(\ref{fp1_122}) and~(\ref{fp2_est}), we get that
\[\frac{\#{\cal F}_n}{\#{\cal E}_n(t_n)} \geq 1 - \frac{\#{\cal F}(p1) + \#{\cal F}(p2)}{\#{\cal E}_n(t_n)} \geq 1-\frac{42}{\omega_n},\]
proving~(\ref{frac_pattern}).~\(\qed\)


\setcounter{equation}{0}
\renewcommand\theequation{\thesection.\arabic{equation}}
\section{Proof of redundancy lower bounds} \label{pf_red_min}
We use the following standard deviation result.
Let~\(\{X_j\}_{1 \leq j \leq m}\) be independent Bernoulli random variables with~\[\mathbb{P}(X_j = 1) = p_j = 1-\mathbb{P}(X_j = 0)\] and
fix~\(0 < \epsilon  \leq \frac{1}{2}.\) If~\(T_m = \sum_{j=1}^{m} X_j\) and~\(\mu_m = \mathbb{E}T_m,\) then
\begin{equation}\label{conc_est_f}
\mathbb{P}\left(\left|T_m - \mu_m\right| \geq \mu_m \epsilon \right) \leq 2\exp\left(-\frac{\epsilon^2}{4}\mu_m\right)
\end{equation}
for all \(m \geq 1.\)\\
\emph{Proof of~(\ref{conc_est_f})}: From Corollary A.1.14, pp. 312 of Alon and Spencer (2008),
\begin{equation}\label{conc_est_f2}
\mathbb{P}\left(\left|T_m - \mu_m\right| \geq \mu_m \epsilon \right) \leq 2\exp\left(-C(\epsilon)\mu_m\right)
\end{equation}
where
\[C(\epsilon) = \min\left(\frac{\epsilon^2}{2}, -\epsilon + (1+\epsilon)\log(1+\epsilon)\right).\]

For any~\(\epsilon > 0,\)~\(\log(1+\epsilon) > \epsilon - \frac{\epsilon^2}{2}\)
and so in particular for~\(0 < \epsilon \leq \frac{1}{2},\)
~\[-\epsilon + (1+\epsilon)\log(1+\epsilon) \geq -\epsilon + (1+\epsilon)\left(\epsilon - \frac{\epsilon^2}{2}\right) = \frac{\epsilon^2}{2} - \frac{\epsilon^3}{2} \geq \frac{\epsilon^2}{4}.\] Thus~\(C(\epsilon) \geq \frac{\epsilon^2}{4}.\)~\(\qed\)

We use a preliminary result regarding number of vectors with long runs.
Let~\(\mathbf{x} = (x_1,\ldots,x_n) \in \{0,1\}^{n}\) be any vector.
For integers~\(i,r \geq 1,\) the vector
\begin{equation}\label{sec_vec}
\mathbf{x}(i,i+r-1) := (x_{i},x_{i+1},\ldots,x_{i+r-1})
\end{equation}
is said to be a run of length~\(r\) if~\(x_{j} = x_{k}\) for all~\(i \leq j,k \leq i+r-1\) and~\(x_{i-1} \neq x_{i}\)
and~\(x_{i+r} \neq x_{i}.\) Let~\(Q_{b+1}(\mathbf{x})\) be the number of runs
of length~\(b+1\) or more in~\(\mathbf{x}\) and let
\begin{equation}\label{e_def}
{\cal U}_n = {\cal U}_n(b) := \left\{\mathbf{x} \in \{0,1\}^{n} : Q_{b+1}(\mathbf{x}) \geq \frac{n}{(b+4)2^{b+4}}\right\}.
\end{equation}
We have the following result.
\begin{Lemma}\label{run_lem} There are positive constants~\(K_i = K_i(b), i =1,2\) such that
for all integers~\(n \geq K_1,\)
\begin{equation}\label{e_est}
\#{\cal U}_n = \#{\cal U}_n (b)  \geq 2^{n}(1-e^{-K_2 n}).
\end{equation}
\end{Lemma}
\emph{Proof of Lemma~\ref{run_lem}}: Let~\(\mathbf{X} = (X_1,\ldots,X_n)\) be a uniformly randomly chosen word in~\(\{0,1\}^{n}\)
so that~\(X_i, 1 \leq i \leq n\) are independent and identically distributed (i.i.d.)
with~\(\mathbb{P}(X_1 = 0) = \frac{1}{2} = \mathbb{P}(X_1 = 1).\)

To lower bound~\(Q_{b+1}(\mathbf{X}),\) we proceed as follows.
Let~\(W\) be the largest integer~\(x\) such that~\(x\cdot (b+3) \leq n\) so that
\begin{equation}\label{w_def}
\frac{n}{b+3} - 1 \leq W \leq \frac{n}{b+3}.
\end{equation}
Define
\begin{equation}\label{i1_def}
I_1 := \{X_1 = 1\} \bigcap_{i=2}^{b+2} \{X_i = 0\} \bigcap \{X_{b+3} = 1\}
\end{equation}
to be the event that~\(\mathbf{X}(1,b+3)\) contains a run of length~\(b+1\) consisting of zeros.
Similarly, for~\(2 \leq r \leq W,\) let~\(I_r\) be the event that the block\\\(\mathbf{X}((r-1)(b+3)+1,r(b+3))\) contains a run of consecutive
zeros of length~\(b+1.\)

The events~\(\{I_i\}_{1 \leq i \leq W}\) are i.i.d. with
\(\mathbb{P}(I_r) = \left(\frac{1}{2}\right)^{b+3}\) and
\begin{equation}\label{q_low}
Q_{b+1}(\mathbf{X}) \geq \sum_{i=1}^{W} \ind(I_i).
\end{equation}
From the deviation estimate~(\ref{conc_est_f}), there is a positive constant~\(C_1\) such that
\[\mathbb{P}\left(\sum_{i=1}^{W} I_i \geq \frac{W}{2^{b+4}}\right) \geq 1-e^{-C_1W}\] and so
using~(\ref{w_def}) we get
\begin{equation}\label{w_est1}
\mathbb{P}\left(\sum_{i=1}^{W} I_i \geq \left(\frac{n}{b+3}-1\right)\frac{1}{2^{b+4}}\right) \geq 1-e^{-C_2n},
\end{equation}
for some constant~\(C_2 > 0.\) Since~\(\frac{n}{b+3}-1 \geq \frac{n}{(b+4)}\) for all~\(n\) large,
we therefore get from~(\ref{q_low}) that
\begin{equation}\label{w_est2}
\mathbb{P}\left(Q_{b+1}(\mathbf{X}) \geq \frac{n}{(b+4)2^{b+4}}\right) \geq 1-e^{-C_2n}.
\end{equation}
This proves~(\ref{e_est}).~\(\qed\)

\emph{Proof of~(\ref{red_tn_conv}) in Theorem~\ref{prop1}}:
Let~\({\cal D}\) be any code capable of correcting up to~\(t_n\) deletable errors.
The code~\({\cal D}\) is capable of correcting up to~\(t_n\) deletions and
we therefore henceforth consider only deletions.
Let~\({\cal F}_r\) be the set of all possible error patterns containing exactly~\(r\) deletions
and no other deletable errors and let~\({\cal F} = \cup_{0 \leq r \leq t_n} {\cal F}_r\)
be the set of all possible error patterns with at most~\(t_n\) deletions.
For~\(\mathbf{x} \in {\cal D},\)
let~\({\cal N}(\mathbf{x}) = \cup_{\mathbf{g} \in {\cal F}}\{F_{\mathbf{g}}(\mathbf{x})\}\) be the set of all vectors in~\(\cup_{0 \leq r \leq t_n}\{0,1\}^{n-r}\) which are obtained from~\(\mathbf{x}\) after corruption by at most~\(t_n\) deletions.

By definition, if~\(\mathbf{x}_1 \neq \mathbf{x}_2 \in {\cal D},\) then necessarily
\[{\cal N}(\mathbf{x}_1) \bigcap {\cal N}(\mathbf{x}_2) = \emptyset,\] because
otherwise~\({\cal D}\) would not be capable of correcting up to~\(t_n\) deletions
and therefore
\begin{equation}\label{sum_less11}
\sum_{\mathbf{x} \in {\cal D}} \#{\cal N}(\mathbf{x}) \leq \sum_{r=0}^{t_n} 2^{n-r} \leq 2^{n+1}.
\end{equation}
Letting~\({\cal U}_n = {\cal U}_n(3)\) be the set as defined in~(\ref{e_def}) with~\(b = 3,\) we obtain a lower bound
on~\(\#{\cal N}(\mathbf{x})\) for each word~\(\mathbf{x} \in {\cal D} \bigcap {\cal U}_n.\)

If~\(\mathbf{x} \in {\cal U}_n,\) then there are at least~\(\frac{n}{7\cdot 2^{7}} \geq \frac{n}{2^{10}} =: \delta n\) runs in~\(\mathbf{x},\)
each of length at least~\(3.\) Choosing~\(t_n\) such runs and deleting one bit in such run, we get a set of distinct corrupted
words. Since there at least~\({\delta n \choose t_n}\) ways to choose a run, we have that~\({\cal N}(\mathbf{x}) \geq {\delta n \choose t_n}.\)

From~(\ref{sum_less11}) we therefore get
\begin{equation}\nonumber
{\delta n \choose t_n} \cdot \#\left({\cal D} \bigcap {\cal U}_n\right) \leq \sum_{\mathbf{x} \in {\cal D} \bigcap {\cal U}_n} {\cal N}(\mathbf{x}) \leq 2^{n+1}
\end{equation}
and so
\begin{equation}\label{dd_er}
\#{\cal D} \leq \frac{2^{n+1}}{{\delta n \choose t_n}} + \#{\cal U}^c_n \leq \frac{2^{n+1}}{2^{\Delta}} + 2^{n} e^{-K_2 n},
\end{equation}
where~\(2^{\Delta} = {\delta n \choose t_n}\) and the final estimate in~(\ref{dd_er}) follows from~(\ref{e_est}).

To show that~\(2^{\Delta}\) is much smaller than~\(e^{K_2 n},\) we set~\(k = \delta n, r = t_n\) and use Stirling's formula~\(r! \geq C_1^{-1} r^{r}e^{-r} \sqrt{r}\) for some constant~\(C_1 > 0\) to get
\begin{equation}\label{kr_est}
{k \choose r} = \frac{k(k-1)\ldots(k-r+1)}{r!} \leq C_1\frac{k^{r}}{r^{r}e^{-r}\sqrt{r}} \leq C_1\left(\frac{ke}{r}\right)^{r}.
\end{equation}
Thus \(2^{\Delta} \leq C_1 \left(\frac{\delta n e}{t_n}\right)^{t_n}\)
and so \[\Delta \leq t_n \log\left(\frac{\delta n e}{t_n}\right) + \log{C_1} \leq t_n\log{n} + \log{C_1},\]
which is much smaller than~\(K_2 n\) since~\(\frac{t_n \log{n}}{n} \longrightarrow 0\) as~\(n \rightarrow \infty.\)

From~(\ref{dd_er}), we therefore get that~\(\#{\cal D} \leq \frac{2^{n+1}}{2^{\Delta}}\) and so the redundancy~\(R({\cal D}) \geq \Delta - 1\)
for all~\(n\) large. To find a lower bound for~\(\Delta,\) we again set~\(k=\delta n, r =t_n\) and use
\[{k \choose r} = \frac{k(k-1)\ldots (k-r+1)}{r!} \geq \frac{(k-r)^r}{r^r}\] to get
\begin{equation}\label{del_tm1}
\Delta \geq t_n \log\left(\frac{\delta n}{t_n} - 1\right) = t_n \log\left(\frac{n}{t_n}\right) -10 t_n + t_n\log\left(1-\frac{t_n}{\delta n}\right),
\end{equation}
since~\(\delta = \frac{1}{2^{10}}.\) But~\(\frac{t_n}{\delta n} \longrightarrow 0\) as~\(n \rightarrow \infty\) and so we have that
\[\left|\log\left(1-\frac{t_n}{\delta n}\right)\right| \leq \frac{2t_n}{\delta n}\] for all~\(n\) large and therefore
\[\Delta \geq t_n \log\left(\frac{n}{t_n}\right) -10 t_n - \frac{2t_n^2}{\delta n} = t_n\left(\log\left(\frac{n}{t_n}\right) - 10 - \frac{2t_n}{\delta n}\right),\] which is positive for all~\(n\) large,
since~\(\frac{t_n\log{n}}{n} \longrightarrow 0\) and so~\(\frac{n}{t_n} \longrightarrow \infty\)\\as~\(n \rightarrow \infty.\)~\(\qed\)

\emph{Proof of~(\ref{min_red}) in Theorem~\ref{thm3}}:
Let~\({\cal D}\) be any code capable of correcting any deletable error burst of length~\(b.\)
As before we consider only deletions since the code~\({\cal D}\) is capable of correcting a burst
deletable error of length~\(b\) consisting only of deletions.
Let~\({\cal F}\) be the set of all burst deletable error patterns of length~\(b\) satisfying~(\ref{err_bur})
and consisting only of deletions. For~\(\mathbf{x} \in {\cal D},\)
let~\({\cal N}(\mathbf{x}) = \cup_{\mathbf{e} \in {\cal F}}\{F_{\mathbf{e}}(\mathbf{x})\}\) be the set of all vectors in~\(\cup_{0 \leq r \leq b} \{0,1,\varepsilon\}^{n-r}\) which are obtained from~\(\mathbf{x}\) after corruption by a burst deletion of length~\(b.\)

By definition, if~\(\mathbf{x}_1 \neq \mathbf{x}_2 \in {\cal D},\) then necessarily
\[{\cal N}(\mathbf{x}_1) \bigcap {\cal N}(\mathbf{x}_2) = \emptyset,\] because
otherwise~\({\cal D}\) would not be capable of correcting a burst deletion of length~\(b.\)
Therefore
\begin{equation}\label{sum_less1}
\sum_{\mathbf{x} \in {\cal D}} \#{\cal N}(\mathbf{x}) \leq \sum_{r=1}^{b} 2^{n-r} \leq b \cdot 2^{n}.
\end{equation}
Letting~\({\cal U}_n\) be the set as defined in~(\ref{e_def}) we obtain a lower bound
on~\(\#{\cal N}(\mathbf{x})\) for each word~\(\mathbf{x} \in {\cal D} \bigcap {\cal U}_n\)
defined in~(\ref{en_def}).

If~\(\mathbf{x} \in {\cal U}_n,\) then there are~\(k \geq \frac{n}{(b+4)2^{b+4}}\) runs in~\(\mathbf{x},\)
each of length at least~\(b+1.\) Let~\(\mathbf{x}(i_1,j_1),\mathbf{x}(i_2,j_2),\ldots,\mathbf{x}(i_k,j_k)\)
be the runs in~\(\mathbf{x}\) so that \[i_1 < j_1 < i_2 < j_2 <\ldots <j_k \text{ and }j_u-i_u \geq b+1\] for all~\(1 \leq u \leq k.\)
If a burst deletion pattern~\(\mathbf{e}\) of length~\(b\) deletes only bits of~\(\mathbf{x}\) with indices between~\(i_1\) and~\(j_1\)
and another burst deletion pattern~\(\mathbf{f}\) of length~\(b\) deletes only bits with indices between~\(i_2\) and~\(j_2\)
then the two resulting deleted words~\(F_{\mathbf{e}}(\mathbf{x}) \neq F_{\mathbf{f}}(\mathbf{x}).\)
Since there are~\(k \geq \frac{n}{(b+4)2^{b+4}}\) ways to choose a run, we have that~\({\cal N}(\mathbf{x}) \geq \frac{n}{(b+4)2^{b+4}}.\)

From~(\ref{sum_less1}) we therefore get
\begin{equation}\nonumber
\frac{n}{(b+4)2^{b+4}} \cdot \#\left({\cal D} \bigcap {\cal U}_n\right) \leq \sum_{\mathbf{x} \in {\cal D} \bigcap {\cal U}_n} {\cal N}(\mathbf{x}) \leq b \cdot 2^{n}
\end{equation}
and so
\begin{equation}\nonumber
\#{\cal D} \leq b(b+4)2^{b+4} \cdot \frac{2^{n}}{n} + \#{\cal U}^c_n \leq b(b+4)2^{b+4} \cdot \frac{2^{n}}{n} + 2^{n} e^{-K_2 n} \leq b(b+4)2^{b+5} \frac{ 2^{n}}{n},
\end{equation}
for all~\(n\) large, using~(\ref{e_est}). This proves~(\ref{min_red}).~\(\qed\)

To prove~(\ref{min_red2}) in Theorem~\ref{thm_11}, we need a small preliminary estimate. For~\(\mathbf{x} \in \{0,1\}^{n}\)
and integer~\(r \geq 1,\)
let
\begin{equation}\label{xr_def2}
\mathbf{x}_r := \mathbf{x}_r(\mathbf{x}) = \mathbf{x}((3P+5)(r-1)+1,(3P+5)(r-1)+5)
\end{equation}
be a sequence of five bits of~\(\mathbf{x}.\) Say that~\(\mathbf{x}_r\) is a good~\(5-\)block if the
first and last bits are one and the remaining three bits are zero.
Let~\(Q(\mathbf{x})\) be the number of good~\(5-\)blocks in~\(\{\mathbf{x}_r\}_{1 \leq r \leq W}\)
where~\(W\) is the largest integer~\(x\) satisfying~\(x\cdot(3P+5) \leq n\) so that
\begin{equation}\label{we_22}
\frac{n}{3P+6} \leq \frac{n}{3P+5} -1 \leq W \leq \frac{n}{3P+5},
\end{equation}
for all~\(n\) large.
Let
\begin{equation}\label{v_def2}
{\cal Y}_n  := \left\{\mathbf{x} \in \{0,1\}^{n} : Q(\mathbf{x}) \geq \frac{n}{64(3P+6)}\right\}.
\end{equation}
\begin{Lemma}\label{run_lem22} We have that
\begin{equation}\label{v_est2}
\#{\cal Y}_n \geq 2^{n}\left(1-\exp\left(-\frac{n}{2^{9}(3P+6)}\right)\right),
\end{equation}
for all~\(n\) large.
\end{Lemma}
\emph{Proof of Lemma~\ref{run_lem22}}: Let~\(\mathbf{X} = (X_1,\ldots,X_n)\) be a uniformly randomly chosen word in~\(\{0,1\}^{n}\)
so that~\(X_i, 1 \leq i \leq n\) are independent and identically distributed (i.i.d.)
with~\(\mathbb{P}(X_1 = 0) = \frac{1}{2} = \mathbb{P}(X_1 = 1).\)

To lower bound~\(Q(\mathbf{X}),\) let~\(r \geq 1\) be an integer and define
\begin{equation}\label{i1_def22}
I_r := \{X_{(3P+5)(r-1)+1} = 1\} \bigcap_{i=2}^{4} \{X_{(3P+5)(r-1)+i} = 0\} \bigcap \{X_{(3P+5)(r-1)+5} = 1\}
\end{equation}
be the event that~\(\mathbf{X}_r\) defined in~(\ref{xr_def2}) is a good block.
The events~\(\{I_r\}\) are i.i.d with~\(\mathbf{P}(I_1) = \frac{1}{32}\)
and so~\(Q(\mathbf{X}) = \sum_{r=1}^{W}I_r\) where~\(W\) is as in~(\ref{we_2})
has mean~\(\frac{W}{32}.\) Therefore using the deviation estimate~(\ref{conc_est_f}) with~\(\epsilon = \frac{1}{2}\)
we get
\begin{equation}\nonumber
\mathbb{P}\left(Q\left(\mathbf{X}\right) \geq \frac{W}{64}\right) \geq 1-\exp\left(-\frac{W}{2^{9}}\right).
\end{equation}
Using~(\ref{we_22}) we have that~\(\frac{W}{64} \geq \frac{n}{64(3P+6)}\) for all~\(n\) large and so
\begin{equation}\nonumber
\mathbb{P}\left(Q\left(\mathbf{X}\right) \geq \frac{n}{64(3P+6)}\right) \geq 1-\exp\left(-\frac{n}{2^{9}(3P+6)}\right)
\end{equation}
for all~\(n\) large.~\(\qed\)

\emph{Proof of~(\ref{min_red2}) in Theorem~\ref{thm_11}}:
Let~\({\cal Y}_n\) be as defined in~(\ref{v_def}) and~\({\cal D}\) be any code capable of correcting all~\(3P-\)far deletable error patterns.
The code~\({\cal D}\) is capable of correcting all~\(3P-\)far deletable error patterns consisting only of deletions
and so we henceforth consider only deletions.

Let~\({\cal F}\) be the set
of all~\(3P-\)far deletable error patters consisting only of deletions  and for~\(\mathbf{x} \in {\cal D},\) set~\({\cal N}(\mathbf{x}) = \cup_{\mathbf{g} \in {\cal F}} F_{\mathbf{g}}(\mathbf{x}).\) Since the deletable errors are at least~\(3P\) apart, there are at most~\(\frac{n}{3P}\) deletable errors in~\(\mathbf{x}\) and so if~\(T\) is the largest integer less than or equal to~\(\frac{n}{3P},\) then~\({\cal N}(\mathbf{x}) \subseteq \cup_{0 \leq r \leq T} \{0,1\}^{n-r}.\) Moreover, if~\(\mathbf{x}_1 \neq \mathbf{x}_2 \in {\cal D},\) then~\(F_{\mathbf{e}}(\mathbf{x}_1) \neq F_{\mathbf{e}}(\mathbf{x}_2)\) by definition. So the sets~\(\{{\cal N}(\mathbf{x})\}\) are all disjoint and therefore
\begin{equation}\label{sum_less322}
\sum_{\mathbf{x} \in {\cal D}} \#{\cal N}(\mathbf{x}) \leq \sum_{r=0}^{T} 2^{n-r} = 2^{n+1}\left(1-\frac{1}{2^{T+1}}\right) \leq 2^{n+1}.
\end{equation}

We now estimate the size of~\({\cal N}(\mathbf{x})\) for each~\(\mathbf{x} \in {\cal D} \bigcap {\cal Y}_n.\)
Since~\(\mathbf{x} \in {\cal Y}_n,\) there are at least~\(\frac{n}{64(3P+6)}\) good~\(5-\)blocks in~\(\mathbf{x}.\) Let~\(m_P\) be the largest integer less than or equal to~\(\frac{n}{64(3P+6)}.\) For every one of the~\(m_P\) good~\(5-\)blocks, we can either choose to remove a bit or not. There are~\(2^{m_P}\) ways of performing such a procedure and the resulting
set of corrupted words are all distinct. Therefore~\(\#{\cal N}(\mathbf{x}) \geq 2^{m_P}\)
and using~(\ref{sum_less322}) we get
\begin{equation}\nonumber
2^{m_P}\#\left({\cal D} \bigcap  {\cal Y}_n \right) \leq \sum_{\mathbf{x} \in {\cal D} \bigcap {\cal Y}_n} \#{\cal N}(\mathbf{x}) \leq 2^{n+1}.
\end{equation}
From~(\ref{v_est2}) we therefore get
\begin{equation}\nonumber
\#{\cal D} \leq \frac{2^{n+1}}{2^{m_P}} + \#{\cal Y}^c_n \leq \frac{2^{n+1}}{2^{m_P}} + 2^{n}\exp\left(-\frac{n}{2^{9}(3P+6)}\right) \leq \frac{2^{n+2}}{2^{\Delta}}
\end{equation}
for all~\(n\) large, where~\(\Delta = \frac{1}{2^{11}(3P+6)}.\) Therefore the redundancy of~\({\cal D}\) is at least~\(\Delta -\log{4} \geq \frac{n}{2^{11}(3P+6)}-2\) for all~\(n\) large. 

To prove~(\ref{min_red22}) in Theorem~\ref{thm_11}, we again need a preliminary estimate. Let~\(\mathbf{x} \in \{0,1\}^{n}\)
and let~\(W\) be the largest integer~\(x\) such that~\(x\cdot 3P_n \leq n\) so that
\begin{equation}\label{we_2}
\frac{n}{3P_n}-1 \leq W \leq \frac{n}{3P_n}.
\end{equation}
For integer~\(1 \leq r \leq W,\) let
\begin{equation}\label{xr_def}
\mathbf{x}_r := \mathbf{x}_r(\mathbf{x}) = \mathbf{x}(3P_n(r-1)+1,3rP_n)
\end{equation}
be a sequence of~\(3P_n\) bits of~\(\mathbf{x}.\) We have thus divided the first~\(3WP_n\) bits of~\(\mathbf{x}\) into~\(W\) disjoint blocks~\(\{\mathbf{x}_r\}_{1 \leq r \leq W},\) each containing~\(3P_n\) bits.

Let~\(U\) be the largest integer~\(y\) such that~\(5 \cdot y \leq 3P_n\)
so that~\(\frac{3P_n}{5}-1 \leq U \leq \frac{3P_n}{5}.\) Divide the first~\(5U\) bits of each block~\(\mathbf{x}_r\) into disjoint blocks~\(\{\mathbf{x}_{r,j}\}_{1 \leq j \leq U}\) of five bits each. Say that~\(\mathbf{x}_{r,j}\) is a good~\(5-\)block if the first and last bits are one and the remaining three bits are zero.
Let~\(T_{r}(\mathbf{x})\) be number of good~\(5-\)blocks in~\(\mathbf{x}_r\) and  let
\begin{equation}\label{v_def}
{\cal V}_n := \left\{\mathbf{x} \in \{0,1\}^{n} : T_r(\mathbf{x}) \geq \frac{3P_n}{64} \text{ for all } 1 \leq r \leq W\right\}.
\end{equation}
Arguing as in the proof of Lemma~\ref{run_lem}, we have the following.
\begin{Lemma}\label{run_lem2} There are positive constants~\(\beta_i,i=1,2\) such that
for all integers~\(n \geq \beta_1,\)
\begin{equation}\label{v_est}
\#{\cal V}_n \geq 2^{n}(1-e^{-\beta_2 P_n}).
\end{equation}
\end{Lemma}
\emph{Proof of Lemma~\ref{run_lem2}}: Let~\(\mathbf{X} = (X_1,\ldots,X_n)\) be a uniformly randomly chosen word in~\(\{0,1\}^{n}\)
so that~\(X_i, 1 \leq i \leq n\) are independent and identically distributed (i.i.d.)
with~\(\mathbb{P}(X_1 = 0) = \frac{1}{2} = \mathbb{P}(X_1 = 1).\)

To lower bound~\(T_r(\mathbf{X})\) for~\(1 \leq r \leq W,\) let~\(I_{r,j}\) be the event that~\(\mathbf{X}_{r,j}\) is a good~\(5-\)block.
The events~\(\{I_{r,j}\}_{1 \leq j \leq U}\) are i.i.d with~\(\mathbf{P}(I_{r,j}) = \frac{1}{32}\)
and so~\(T_r(\mathbf{X}) = \sum_{j=1}^{U} I_{r,j}\) has mean~\(\frac{U}{32}.\) Therefore from the deviation estimate~(\ref{conc_est_f}),
we get that
\begin{equation}\label{q_dev1}
\mathbb{P}\left(T_r\left(\mathbf{X}\right) \geq \frac{U}{64}\right) \geq 1-e^{-C_1U}
\end{equation}
for all~\(n\) large and some constant~\(C_1 > 0.\) Since~\(U \geq \frac{3P_n}{5} -1 \geq \frac{3P_n}{8}\) for all large~\(n,\)
we get
\begin{equation}\label{q_dev2}
\mathbb{P}\left(T_r\left(\mathbf{X}\right) \geq \frac{3P_n}{2^{9}}\right) \geq 1-e^{-4C_2P_n}
\end{equation}
for some constant~\(C_2 >0\) and all large~\(n.\) The constant~\(C_2\) does not depend on~\(r\) and so
\begin{equation}\label{q_dev3}
\mathbb{P}\left(\bigcap_{1 \leq r \leq W} \left\{T_r\left(\mathbf{X}\right) \geq \frac{3P_n}{2^{9}}\right\}\right) \geq 1-W\cdot e^{-4C_2P_n}.
\end{equation}
Using~(\ref{we_2}) and~(\ref{pn_cond}) we get that~\[W\cdot e^{-4C_2P_n} \leq \frac{n}{2P_n} \cdot e^{-4C_2 P_n} \leq e^{-2C_2 P_n}\] for all~\(n\) large.~\(\qed\)

\emph{Proof of~(\ref{min_red2}) in Theorem~\ref{thm_11}}: Let~\({\cal D}\) be any code capable of correcting all~\(3P_n-\)far deletable error patterns.
The code~\({\cal D}\) is capable of correcting all~\(3P_n-\)far deletable error patterns consisting only of deletions
and so we henceforth consider only deletions.
Let~\({\cal F}\) be the set of all~\(3P_n-\)far deletable error patters consisting only of deletions and for~\(\mathbf{x} \in {\cal D},\) set~\({\cal N}(\mathbf{x}) = \cup_{\mathbf{e} \in {\cal F}} F_{\mathbf{e}}(\mathbf{x}).\) Since the deletions are at least~\(3P_n\) apart, there are at most~\(\frac{n}{3P_n}\) deletable errors in~\(\mathbf{x}\) and so if~\(L\) is the largest integer less than or equal to~\(\frac{n}{3P_n},\) then~\({\cal N}(\mathbf{x}) \subseteq \cup_{0 \leq r \leq L} \{0,1\}^{n-r}.\) Moreover, if~\(\mathbf{x}_1 \neq \mathbf{x}_2 \in {\cal D},\) then~\(F_{\mathbf{e}}(\mathbf{x}_1) \neq F_{\mathbf{e}}(\mathbf{x}_2)\) by definition. So the sets~\(\{{\cal N}(\mathbf{x})\}\) are all disjoint and therefore
\begin{equation}\label{sum_less3}
\sum_{\mathbf{x} \in {\cal D}} \#{\cal N}(\mathbf{x}) \leq \sum_{r=0}^{L} 2^{n-r} = 2^{n+1}\left(1-\frac{1}{2^{L+1}}\right) \leq 2^{n+1}.
\end{equation}

We now estimate the size of~\({\cal N}(\mathbf{x})\) for each~\(\mathbf{x} \in {\cal D} \bigcap {\cal V}_n.\)
Let~\(W\) be as in~(\ref{we_2}) and as before, split the first~\(3WP_n\) bits of~\(\mathbf{x}\) into~\(W\) blocks~\(\{\mathbf{x}_r\}_{1 \leq r \leq W}\) of length~\(3P_n\) each. Since~\(\mathbf{x} \in {\cal V}_n,\) there at least~\(\frac{3P_n}{64}\) good~\(5-\)blocks in each~\(\mathbf{x}_r.\)
Choosing one good~\(5-\)block from each of~\(\mathbf{x}_1,\mathbf{x}_3,\ldots\) and deleting a middle zero from each such good block, we get a set of distinct words. Since there are at least~\(\frac{W-1}{2}\) blocks in~\(\mathbf{x}_1,\mathbf{x}_3,\ldots,\) the number of corrupted words obtained as above is at least~\(\left(\frac{3P_n}{64}\right)^{(W-1)/2}.\) The deletable error patterns giving rise to these words are all~\(3P_n-\)far apart and so we get
\begin{equation}\label{nx_f}
\#{\cal N}(\mathbf{x}) \geq \left(\frac{3P_n}{64}\right)^{(W-1)/2} \geq \left(\frac{3P_n}{64}\right)^{\frac{n}{6P_n}-1},
\end{equation}
since~\(W \geq \frac{n}{3P_n}-1,\) (see~(\ref{we_2})).

Letting~\[\Delta = \left(\frac{n}{6P_n}-1\right)\log\left(\frac{3P_n}{64}\right)\] and using~(\ref{nx_f}) in~(\ref{sum_less3}) we get
\begin{equation}\nonumber
2^{\Delta}\#\left({\cal D} \bigcap  {\cal V}_n \right) \leq \sum_{\mathbf{x} \in {\cal C} \bigcap {\cal V}_n} \#{\cal N}(\mathbf{x}) \leq 2^{n+1}
\end{equation}
and so from~(\ref{v_est}) we get
\begin{equation}\nonumber
\#{\cal D} \leq \frac{2^{n+1}}{2^{\Delta}} + \#{\cal V}^c_n \leq \frac{2^{n+1}}{2^{\Delta}} + 2^{n}e^{-\beta_2 P_n} \leq \frac{2^{n+2}}{2^{\Delta}}
\end{equation}
for all~\(n\) large, since using the condition~(\ref{pn_cond}) that~\(\frac{P_n^2}{n\log{n}} \longrightarrow \infty\) as~\(n \rightarrow \infty,\) we have
\[\Delta \leq \frac{n}{P_n}\log(3P_n) \leq \frac{n \log{n}}{P_n} \leq \beta_2 P_n\log{e}\] for all~\(n\) large. Therefore the redundancy of~\({\cal D}\) is at least~\(\Delta -\log{4}\) for all~\(n\) large.~\(\qed\)


\section*{Acknowledgements}
I thank Professors Alberto Gandolfi and Federico Camia for crucial comments and for my fellowships.




\setcounter{equation}{0} \setcounter{Lemma}{0} \renewcommand{\theLemma}{II.%
\arabic{Lemma}} \renewcommand{\theequation}{II.\arabic{equation}} %
\setlength{\parindent}{0pt}




%





\bibliographystyle{plain}

\end{document}